\documentclass{article}
\usepackage{arxiv_style}

\AtBeginDocument{%
  \providecommand\BibTeX{{%
    \normalfont B\kern-0.5em{\scshape i\kern-0.25em b}\kern-0.8em\TeX}}}

\usepackage[utf8]{inputenc} % allow utf-8 input
\usepackage[T1]{fontenc}    % use 8-bit T1 fonts
\usepackage{hyperref} 
\usepackage{scalerel,stackengine}
\usepackage{times}  % DO NOT CHANGE THIS
\usepackage{helvet}  % DO NOT CHANGE THIS
\usepackage{courier}  % DO NOT CHANGE THIS
\usepackage{graphicx} % DO NOT CHANGE THIS
\usepackage[numbers]{natbib}  % DO NOT CHANGE THIS AND DO NOT ADD ANY OPTIONS TO IT
\usepackage{caption} % DO NOT CHANGE THIS AND DO NOT ADD ANY OPTIONS TO IT
\usepackage[many]{tcolorbox}
\usepackage{subcaption}
\colorlet{punct}{red!60!black}
\definecolor{background}{HTML}{EEEEEE}
\definecolor{delim}{RGB}{20,105,176}
\colorlet{numb}{magenta!60!black}
\definecolor{main}{HTML}{5989cf}    % setting main color to be used
\definecolor{sub}{HTML}{cde4ff}     % setting sub color to be used
\tcbset{
    sharp corners,
    colback = white,
    before skip = 0.2cm,    % add extra space before the box
    after skip = 0.5cm      % add extra space after the box
}  

\newtcolorbox{simplebox}{
    fontupper = \bf,
    boxrule = 1.5pt,
    colframe = black, % frame color
    fontupper=\footnotesize,
    fontlower=\footnotesize
}
\newtcolorbox{chatbox1}{
    colback = sub, % background color
    boxrule = 0pt,  % no borders
    fontupper=\footnotesize,
    fontlower=\footnotesize
}
\newtcolorbox{chatbox2}{
    colback = sub, 
    colframe = main, 
    boxrule = 0pt, 
    toprule = 3pt, % top rule weight
    bottomrule = 3pt, % bottom rule weight
    fontupper=\tiny,
    fontlower=\tiny
}

\newcommand{\icwsm}[1]{\textcolor{black}{#1}}
\newcommand{\asset}{\textcolor{cyan}{asset}}
\newcommand{\need}{\textcolor{red}{need}}
\newcommand{\other}{\textcolor{olive}{other}}
\title{Community Needs and Assets: A Computational Analysis of Community Conversations}
\author{
Md Towhidul Absar Chowdhury\\
  \small{Golisano College of Computing and Information Sciences}\\
  \small{Rochester Institute of Technology} \\
  \texttt{mac9908@rit.edu} \\
\And
Naveen Sharma \\
  \small{Golisano College of Computing and Information Sciences}\\
  \small{Rochester Institute of Technology}\\
  \texttt{naveen.sharma@rit.edu} \\
 \And
Ashiqur R. KhudaBukhsh \\
  \small{Golisano College of Computing and Information Sciences}\\
  \small{Rochester Institute of Technology}\\
  \texttt{axkvse@rit.edu} \\
}

\begin{document}

\maketitle
\begin{abstract}
A community needs assessment is a tool used by non-profits and government agencies to quantify the strengths and issues of a community, allowing them to allocate their resources better.
Such approaches are transitioning towards leveraging social media conversations to analyze the needs of communities and the assets already present within them. However, manual analysis of exponentially increasing social media conversations is challenging.
There is a gap in the present literature in computationally analyzing how community members discuss the strengths and needs of the community.
To address this gap, we introduce the task of identifying, extracting, and categorizing community needs and assets from conversational data using sophisticated natural language processing methods.
To facilitate this task, we introduce the first dataset about community needs and assets consisting of 3,511 conversations from Reddit, annotated using crowdsourced workers. \icwsm{Using this dataset, we evaluate an utterance-level classification model compared to sentiment classification and a popular large language model (in a zero-shot setting), where we find that our model outperforms both baselines at an F1 score of 94\% compared to 49\% and 61\% respectively. 
Furthermore, we observe through our study that conversations about needs have negative sentiments and emotions, while conversations about assets focus on location and entities. The dataset is available at \href{https://github.com/towhidabsar/CommunityNeeds}{https://github.com/towhidabsar/CommunityNeeds}\footnote{Accepted at The International AAAI Conference on Web and Social Media (ICWSM) 2024}.}
\end{abstract}

\section{Introduction}
Understanding the needs and assets of neighborhoods is an important task for non-profits, government agencies, and local leaders to affect positive development within their communities. Such development work can be in the form of shelter programs to alleviate homelessness or educational programs for young adults. However, with limited resources at their disposal, organizations have to choose what to prioritize; whether to focus on resolving the unmet needs or whether to foster the assets of the community they are serving.  

\begin{figure}[t]
    \centering
\begin{chatbox1}
\textbf{Post:}

My amazing grandmother lives in Flatbush, Brooklyn, and in desperate need of someone to come help her, a therapist and a cool, artsy, and kind community to be around and hang out with. She is struggling with serious anxiety attacks and keeps ending up at the psych ward....I’m looking for any advice on care for the elderly and mental health in Brooklyn, and/or any charities or communities in the area who might be able to help. if anyone has had any experience with this, I would be so appreciative. 

\textbf{Comment:}

Flatbush seriously lacking a senior center that is desperately needed!!! Contact Crystal Hudson's office, she is Chair of Aging Committee for City Council, dedicated to helping with Seniors [https://council.nyc.gov/district-35/](https://council.nyc.gov/district-35/)

\textbf{Label:} Need - Support for Special Population
\end{chatbox1}
\caption{An example of a conversation about community needs; the need in this example is being ``Support for Special Population''.}
\label{fig:need_example}
\end{figure}

In the non-profit literature, a ``need'' is a discrepancy between ``what is'' and ``what should be'', while an ``asset'' (resource) is anything that can be used to improve the quality of community life~\cite{mcknight_building_1993}. A ``community needs assessment'' is a systematic set of procedures that are used to determine needs and assets, examine their nature and causes, and set priorities for future actions~\cite{witkin_planning_1995}.  Understanding community needs and assets is a vital domain consideration to help non-profits and government agencies perform their activities better. 

Community needs assessments are traditionally done through surveys~\cite{billings_approaches_1995}, focused group discussions~\cite{williams_public_2020}, and manual data analysis by human actors. Manual data analysis includes the synthesis of public open datasets~\cite{al-qdah_syrian_2017} and historical assessment data~\cite{billings_approaches_1995}.

% Importance of conversational lense
The exponential growth of community-related discourses in social media has led to non-profits only recently incorporating social network analysis~\cite{alonzo_using_2023} as a part of their community assessments; using tools such as CrowdTangle\footnote{https://www.crowdtangle.com/} to supplement their traditional methods and study topics of interest in their target communities~\cite{white_mapping_2023}. 
Analyzing community conversations provides an additional valuable lens to established survey-based methods to study a community\icwsm{~\cite{iqbal_lady_2023,lasri_large-scale_2023}} \icwsm{and increases the validity of findings of such assessments as highlighted by \citet{zohrabi_mixed_2013}}.
Another key benefit from traditional methods of needs assessment is that we are not surveying the community and asking them questions, they have free and open discussions from which we are extracting value. 

For example, in Figure~\ref{fig:need_example}, a community resident is looking for support for senior citizens while another resident comments about how the area has a significant lack of such support. Similarly, Figure~\ref{fig:asset_example} includes a comment about the charitable work done by churches in the neighborhood highlighting a key asset in the community. 
Identifying and locating these community conversations from noisy data will provide strong support to established traditional methods of community assessments. However, due to the large amount of data, manually performing this task may overwhelm human actors~\cite{shah_big_2015}. 
Our task is to identify and extract such conversations leveraging established state-of-the-art methods in natural language processing. 
\begin{figure}[t]
    \centering
\begin{chatbox1}
A churches charitable work should absolutely be tax deductible. They absolutely do good work. Everything else however should not. Throwing events for members should be taxed money, so should real estate etc etc. You shouldn't be allowed to donate property to a church then it ``hires'' you and compensates you in letting you live there so you don't pay taxes. There's a million and one tax loopholes involving churches that people exploit.

\textbf{Label:} Asset - Institutional and Civic Asset
\end{chatbox1}
\caption{An example of a conversation about community assets; the asset in this example is ``Institutional and Civic Asset''.}
\label{fig:asset_example}
\end{figure}

Particularly with the widespread adoption of large language models in social science~\cite{ziems_can_2023}, we need to investigate what differentiates the conversations about needs and assets from other conversations. Such techniques have been used to identify community health needs during the COVID-19 pandemic~\cite{sv_analysis_2021,mejova_comfort_2023}, utilizing NLP techniques.
However, the scope of such studies and methods focused on a subset of only needs (health and diet) and does not include linguistic analysis. 
There is a need for a systematic computational approach to understanding and analyzing community needs and assets (CNA) with a robust dataset for evaluating future methods.

In this paper, we ask two research questions: 
\begin{itemize}
    \item (\textbf{RQ1}) Can we computationally extract community needs and asset information from community conversations in social media?
    \item (\textbf{RQ2}) What are the linguistic features that differentiate community needs and assets conversations? 
\end{itemize}

To study our research questions, we model identifying needs and assets as an utterance-level classification task to deconstruct and find linguistic patterns. 
An {\it{utterance}} is a unit of communication produced by a single speaker to convey a single subject, which may span multiple sentences~\cite{bakhtin_speech_2010,ziems_can_2023}. 
Our contributions to this paper are:
\begin{enumerate}
    \item We introduce a novel Community Needs and Assets (CNA) dataset of conversations from \icwsm{eleven} geographical communities on Reddit to identify community needs and assets from natural language conversations. The dataset is annotated by crowd-sourced workers where each conversation is assigned the labels \need, \asset, or \other. It aims to provide a much-needed benchmark for utterance-level classification tasks for mining conversations about community needs and assets in computational social science. 
    \item Using our dataset, we perform baseline computational analysis using supervised classification, zero-shot approaches, and sentiment analysis to evaluate the feasibility of existing approaches that can be used to extract community needs and assets. We find that zero-shot and sentiment analysis approaches perform poorly off-the-shelf in identifying such conversations indicating that there is a need for such dataset to progress towards a more robust computational analysis of community needs assessments.
    \item Using the baseline utterance-level classification model to extract community needs and assets, we take a computational linguistic approach to deconstruct how conversations about community needs and assets are classified by such supervised methods. To our knowledge, our study is the first analysis of needs and assets from a conversational lens that can supplement traditional community needs assessment methods. We find that conversations about \need~are attached to negative sentiment \icwsm{but not all negative conversations are focused on community needs}. Furthermore, \asset~based conversations are not only attached to positive sentiments but are focused more on identifying specific entities and locations. 
    % \textcolor{green}{TODO: Update Introduction after all the changes. Contextualize and strengthen this claim.}

\end{enumerate}
\section{Related Work}\label{sec:related}
% Our work to perform a computational analysis of needs-based discussions using large language models builds on existing multi-disciplinary work in computational social science. We provide the relevant literature and background to both the social science and the computational aspects in this section.
The widespread use of needs assessment originated in the United States with federal government programming in the 1960s. The concept of urban community assessments and needs assessment was formalized in 1995~\cite{witkin_planning_1995,billings_approaches_1995} as a primary method of discerning the gap between what is available and what should be available i.e. ``need'' for a target group (in our case an ``urban community''). 
These assessments have become a prominent tool for data collection with the increase in data-driven decision-making for communities and neighborhoods~\cite{kingsley_strengthening_2014,chowdhury_citizenly_2021}. 
Recent works are transitioning more towards the use of artificial intelligence for such development work~\cite{vinuesa_role_2020} and integrating social media for assessments~\cite{alonzo_using_2023}. \icwsm{Extant research has utilized social media conversations to deconstruct altruistic requests in Stack Overflow~\cite{althoff_how_2014} and linguistic analysis of social comparisons using Twitter~\cite{cui_social_2022}. Our work on identifying the needs and assets of a community from social media conversations can be situated in the same vein as these works. Similar to the linguistic analysis performed by~\citet{giorgi_author_2023} in deconstructing personal narratives, we deconstruct how people identify what is needed and what provides strength to their local community. Furthermore, our work can also be situated alongside recent advances in demographic and geographic inference as well as political analysis of communities on social media~\cite{iqbal_lady_2023,herdagdelen_geography_2023,lasri_large-scale_2023}}

We utilize an utterance-level classification~\cite{ziems_can_2023} approach which include classifying dialects~\cite{demszky_analyzing_2019}, emotions~\cite{ortony_cognitive_2022}, hate speech~\cite{elsherief_latent_2021}, stance~\cite{dutta_murder_2022}, and misinformation detection~\cite{alam_fighting_2021}. We approach community conversations at the utterance level of abstraction \icwsm{and can be situated alongside the recent advances in stance mining but instead of political ideology~\cite{jiang_retweet-bert_2023}, we focus our work on the identification of community needs and assets, as we provide our CNA dataset to evaluate approaches to computational community needs assessments.}

\section{Community Needs and Assets}\label{sec:taxonomy}
\subsection{What are Community Needs?}
Existing needs assessment approaches define community needs depending on their target community \cite{witkin_planning_1995}. For a computational analysis, we categorize needs as defined in the ``Community Needs Assessment'' performed by the New York City Department of Youth and Community Development~\cite{nyc_dycd_community_2022}. 
The categories and sub-categories are as follows: 
\paragraph{Basic Needs} Describes the fundamental necessities in the neighborhood such as food and nutrition assistance, health care, financial assistance, legal services, transportation, crime prevention, etc.
\paragraph{Education} Describes services and programs to help education such as adult education/literacy, college preparation, financial literacy, etc.
\paragraph{Employment} Includes services such as career counseling, assistance starting a business, job skills training, etc.
\paragraph{Out of School Time (School)} DYCD includes afterschool programs and summer recreation services under this umbrella category - differentiating from education with a focus on recreation.
\paragraph{Family Supports} Includes childcare and early childhood development. Additionally, this also covers support for domestic violence victims, family counseling, and parenting support.
\paragraph{Support for Special Population (SP)} This category includes services for senior citizens, veterans, or the disabled.

\subsection{What are Community Assets?}
The Community Capitals Framework \cite{flora_rural_2016} models communities as a system of assets that interact with each other to generate value and capital. Each of these assets is a sub-system of its own~\cite{chowdhury_community_2022} and a taxonomy of the value within the community is already defined by Callaghan \cite{callaghan_building_2008} with four categories of assets:
\paragraph{Human Assets}
Human assets are the skills and abilities of each individual within a community.
Residents who have the ability to build and transform their own community. This includes but is not limited to teachers, community organizers, volunteers, elected officials, and local business owners~\cite{schultz_community_2000}.
\paragraph{Institutional and Civic Assets (IC)}
Community services like public transportation, early childhood education centers, recycling facilities, and cultural organizations improve the lives of community members, whether they operate as nonprofits, for-profits, or government entities. These institutional and civic assets offer programs, services, and commerce opportunities~\cite{schultz_community_2000}.
\paragraph{Physical or Built Assets}
It could be a physical location like a school, hospital, church, library, recreation center, or social club, serving as a town landmark or symbol. It may also include unused buildings or vacant land suitable for a community hospice or meeting room on the second floor. Alternatively, it could be a public space like a park, wetland, or open area already owned by the community~\cite{schultz_community_2000,callaghan_building_2008,flora_rural_2016}.
\paragraph{Cultural Assets}
Cultural assets are the arts, music, language, traditions, stories, and histories that make up a community’s identity, character, and customs. This asset is harder to define as it may contain aspects of a community such as ethnic, racial, or religious diversity. It may also contain concrete things such as historical sites or festivals and fairs~\cite{callaghan_building_2008,flora_rural_2016,schultz_community_2000}.
\section{The CNA Dataset}
\begin{table}[t]
    \centering
    \begin{tabular}{|l r|}
    \hline
    \textbf{Community} & \textbf{Count}\\
    \hline
    \hline
        % Brooklyn   &    712     &   267\\
        % Queens     &    241     &    47\\
        % Rochester  &   1,548     &   601\\
        % bronx      &     84     &    34\\
        % manhattan  &      6     &     4\\
        % nyc        &   4,081     &   1465\\
        % Total        &   6,672     &   2,418\\
        Manhattan       &       4\\
        Bronx           &       33\\
        Queens          &       46\\
        Brooklyn        &       264\\
        NYC             &       1,449\\
        Rochester       &       599\\
        Colorado Springs&       400\\
        Virginia Beach  &       200\\
        Jacksonville    &       200\\
        Mesa, AZ        &       116\\
        Oklahoma City, OK &     200\\
        \hline
        \hline
        Total      &    3,511\\
    \hline
    \end{tabular}
    \caption{\icwsm{The dataset statistics by Reddit communities in our CNA dataset.}}
    \label{tab:data_stat}
\end{table}

Reddit is a social media platform that has been extensively used for several computational social science studies\icwsm{~\cite{giorgi_author_2023,lokala_computational_2022}}. It is known for its forum and discussion-oriented post structure with an emphasis on separated communities (\textit{subreddits}).
This allows conversations to be categorized and targeted to only specific communities and suits our community-based analysis.

\subsection{Dataset Construction} 
\subsubsection{Keyword and Subreddit Selection}
We create a dataset by collecting posts and comments from \icwsm{11} \textit{subreddits} as shown in Table~\ref{tab:data_stat}. We start with a seed set of keywords relevant to the community needs ~(1)~\texttt{community},~(2)~\texttt{community needs},~(3)~\texttt{community school},~(4)~\texttt{programs services household receive}, and~(5)~\texttt{welfare} to ensure initial relevance to community discussions. 
\texttt{programs services household receive} was selected from the questionnaire used by an actual needs assessment~\cite{nyc_dycd_community_2022} to see if such keywords help with finding relevant conversations. Other keywords were experimented with as well but did not add any new conversations in addition to the ones provided by these five keywords.

\icwsm{We selected 11 subreddits, 5 of them representing each of the communities in the New York City area along with the subreddit of \textit{Rochester} to allow for a needs assessment of mid-sized cities within similar geographic locations. We selected the five boroughs of New York and Rochester because the state of New York provides socioeconomic diversity within a shared regional context. By including examples from different boroughs with diverse demographics we can account for differences in needs and assets. As a global city, the population of New York City represents tremendous cultural, ethnic, and linguistic diversity which adds to the generalization of our dataset.}

\icwsm{However, New York City and Rochester combined are still more socially and economically liberal. Focusing only on NYC communities misses potential insights from other urban regions that are more conservative for comparison. 
Needs and assets conversations may differ across geographic and political contexts.
To enable our CNA dataset to be generalizable to urban communities, we further included the five most conservative big cities in the United States in our dataset~\cite{tausanovitch_representation_2014}: (1) \textit{Mesa, AZ}, (2) \textit{Oklahoma City, OK}, (3) \textit{Virginia Beach, VA}, (4) \textit{Colorado Springs, CO}, and (5) \textit{Jacksonville, FL}}

For these~\textit{subreddits}, we retrieved all posts and comments using a widely used library\footnote{https://praw.readthedocs.io} limiting our search results to posts that have at least 5 comments to ensure there was some minimal amount of user engagement.

\subsubsection{Corpus Filtering}\label{sec:corpus_reduction}
We further synthesize our community needs corpus from the initial Reddit corpus with a natural language inference (NLI) approach where we calculate the probability of entailment, contradiction, or neutral relevance between each conversation, premise $P$,  and our hypothesis ``$H$: \textit{Community needs are important}''. This allows our corpus to only contain conversations that are semantically close to the idea of community needs. We adapted this method to compute semantic similarity from existing approaches for aggregated stance mining in computational social science literature~\cite{dutta_murder_2022,halterman_corpus-level_2021,khudabukhsh_fringe_2022,chowdhury2024infrastructure}. We define $NLI(P, H)$ as a function, where $P$ is the premise and $H$ is the hypothesis, yielding an output $o \epsilon {ent, con, neu}$. For a conversation $c_i$ in corpus $C$, we calculate the entailment ratio $ent(c_i, H)$ as the fraction of sentences in $c_i$ that entail the hypothesis $H$.
% We define $NLI(P, H)$ as a function that takes premise $P$ and our hypothesis $H$ as input, and provides an output $o \epsilon \{ent, con, neu\}$ where $ent$ is entailment, $con$ is a contradiction, and $neu$ is neutral. 
% Given corpus $C\epsilon\{c_1, c_2, c_3,...,c_i,..\}$, for a conversation $c_i$ we calculate entailment ratio, $ent(c_i,H)$, as a fraction of the individual sentences in $c_i$ that entails the hypothesis $H$ as follows: 

% $ ent(C_i, H) = \frac{\sum_{P\epsilon C_i} I(NLI(P,H)=ent)}{|C_i|}$

% Here $I$ is the indicator function, and a larger value of $ent(C_i, H)$ indicates greater support of $H$. 
The final corpus $C_{final}$ only contains conversations $c_i$ that have an entailment ratio of over 0.5, resulting in the distribution shown in Table~\ref{tab:data_stat}.
\begin{figure}[h]
    \centering
    \begin{simplebox}
According to the given text, which of the following is this comment talking about?
        \begin{itemize}
        \item Need: A community issue, problem, or need (Something negative the community is concerned with. Or something the community is missing) 
        \item Asset: A community highlight, strength, or asset (Something positive the community has)
        \item Other: Other (If unsure write what it is about)
        \item None: None of the above (the comment is about something else and has nothing to do with the community)
        \end{itemize}
\end{simplebox}
    \caption{Prompt for zero-shot text classification of community needs conversation}
    \label{fig:prompt_class}
\end{figure}
\begin{figure}[h]
    \centering
    \begin{simplebox}
What kind of programs, services, or needs is this comment talking about?
        \begin{itemize}
        \item Basic Needs 
        \item Education 
        \item Employment \& Career Advancement 
        \item Out of School Time
        \item Family Supports
        \item Support for Special Population
        \item Other
        \end{itemize}
\end{simplebox}
    \caption{Question for community needs}
    \label{fig:categorize_needs}
\end{figure}
\begin{figure}[h]
    \centering
    \begin{simplebox}
What kind of highlight, strength, or asset is this comment talking about?
        \begin{itemize}
        \item Human Assets 
        \item Institutional and Civic Assets 
        \item Physical Assets 
        \item Cultural Assets
        \end{itemize}
\end{simplebox}
    \caption{Question for community assets}
    \label{fig:categorize_assets}
\end{figure}

\subsection{Crowd Sourced Annotation}
We annotated each of our comments in $C_{final}$ with crowdsourced workers through an anonymized Amazon MTurk task
\footnote{Task outline: \url{https://osf.io/sydf2/?view_only=b3c0a843e6244f4d99a6b349156adad8}}. 
The primary three questions are given in Figure~\ref{fig:prompt_class}, Figure~\ref{fig:categorize_needs}, and Figure~\ref{fig:categorize_assets}. 

The first question in Figure~\ref{fig:prompt_class} asks the annotator to identify if the conversation is about need, asset, or irrelevant to the community.  
The second question in Figure~\ref{fig:categorize_needs} determines (if the conversation is about a need) what category of need is it about.
The third question in Figure~\ref{fig:categorize_needs} determines (if the conversation is about an asset) what category of asset is it about.
Each comment was annotated by three independent annotators. To ensure the reliability of annotations, we limited annotators to be located within the USA (conversations in our corpus are from communities within the USA) and have MTurk Master's qualification. 

\paragraph{Compensation}
We compensate the annotator 0.15 USD for each instance where each batch with 20 instances would thus fetch 3 USD. 
Compensation is grounded in prior literature as the initial pilot by the authors estimated \$12/hour compensation with a completion time of 15 minutes per task and \$3/task. This is more than the US minimum wage (\$7.25) and falls within the range reported in extant literature (\$6/hour in~\citet{leonardelli_agreeing_2021}; \$7.25/hour in~\citet{bugert_breaking_2020}; and \$13/hour in~\citet{bai_pre-train_2021}).

\begin{figure}
    \centering
    \includegraphics[width=0.8\linewidth]{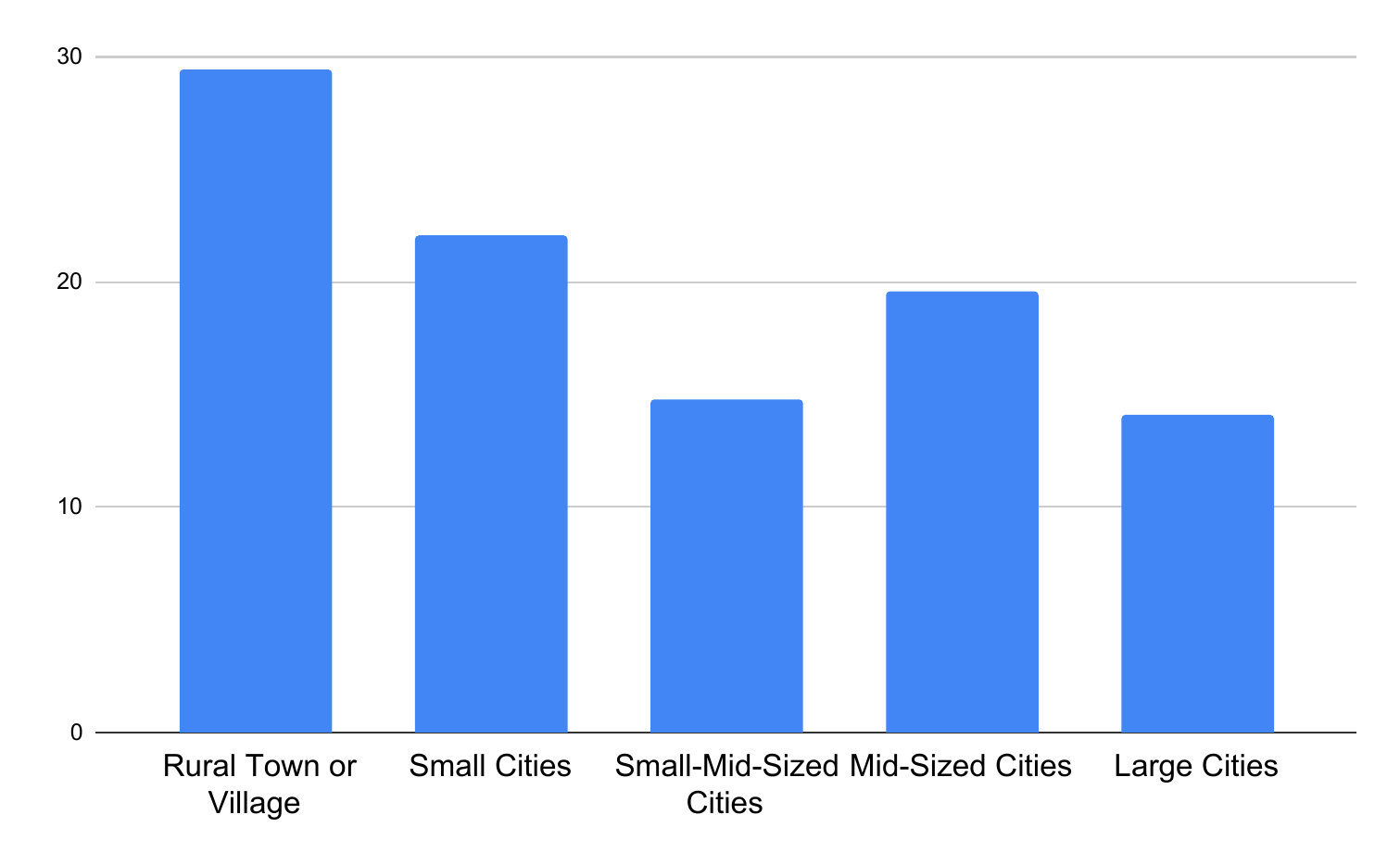}
    \caption{The distribution of community background for the human annotators for CNA dataset. The $y$ axis represents the percentage of annotators.}
    \label{fig:demo_cities}
\end{figure}

\begin{figure}[h]
    \centering
    \includegraphics[width=0.9\linewidth]{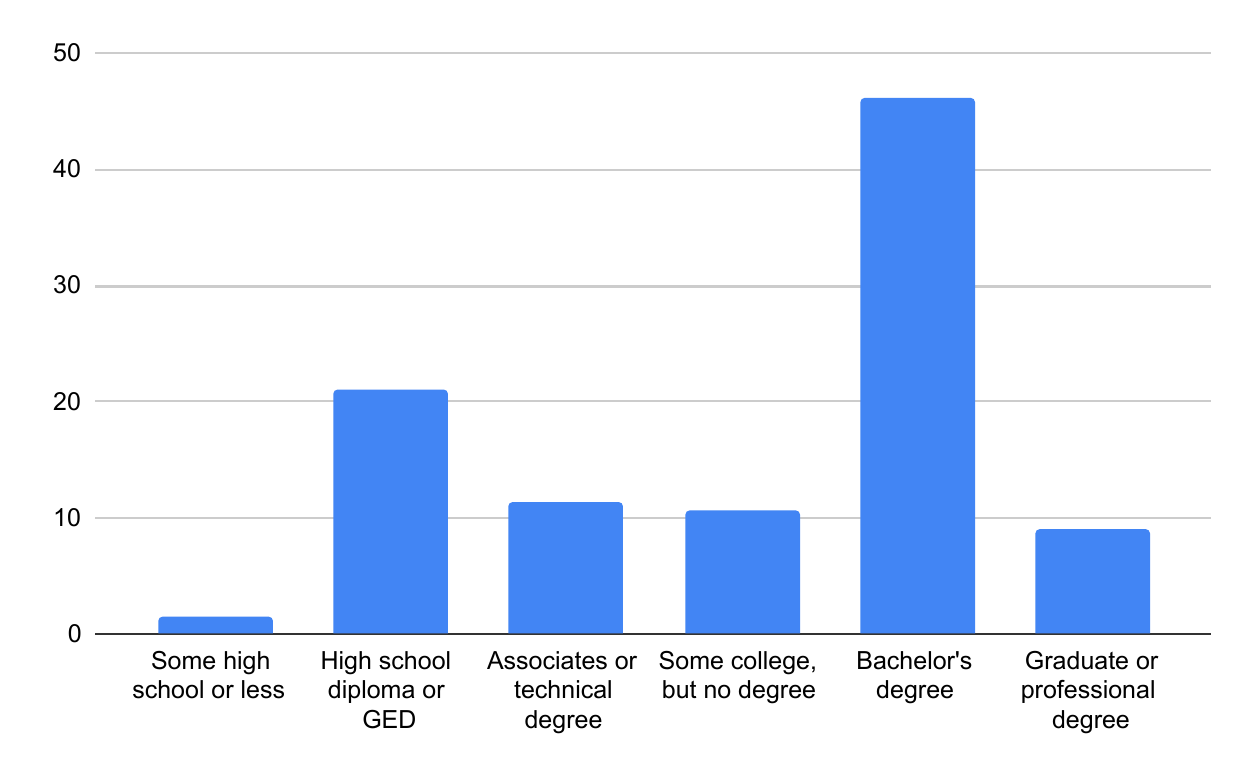}
    \caption{The distribution of educational background for the human annotators for CNA dataset. The $y$ axis represents the percentage of annotators.}
    \label{fig:demo_edu}
\end{figure}

\begin{figure}[h]
    \centering
    \includegraphics[width=0.9\linewidth]{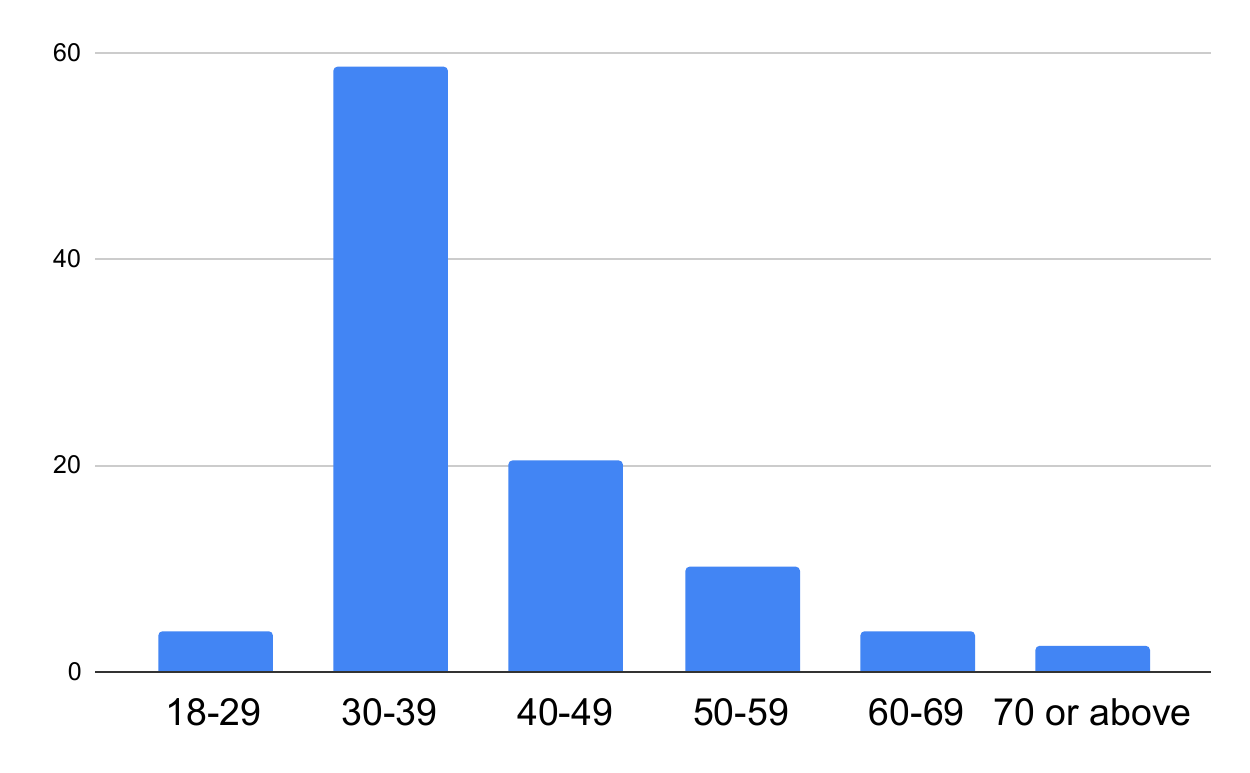}
    \caption{The distribution of age for the human annotators for CNA dataset. The $y$ axis represents the percentage of annotators.}
    \label{fig:demo_ages}
\end{figure}

\paragraph{Demographic}
Approximately 50\% of our annotators are from small cities and rural towns, while the remaining annotators combined come from larger than mid-sized cities as shown in Figure~\ref{fig:demo_cities}. The majority of our annotators are within the age range of $30-39$ (\~60\%) as shown in Figure~\ref{fig:demo_ages} while approximately 45\% of our annotators have a Bachelor's degree (Figure~\ref{fig:demo_edu}). 
\begin{figure}[t]
    \centering
\begin{chatbox1}
\textcolor{red}{This. No one wants to live in a dump but paying \$2500 a month is a pox on anyone's ability build a solid financial future. I know many homeowners in the midwest that don't even pay that on their home mortgages. I understand that NYC is far above its darkest days but it's still grimy, the infrastructure isn't in great shape and has a very deep income inequality.} I think it's hard to justify living in this city when it's also one of the most expensive places to rent/own in the country. \textcolor{blue}{But, the culture, food, and career opportunities are second to none. People love this city, they just wish it wasn't so damn expensive.}
\end{chatbox1}
\caption{\icwsm{An example of a conversation that had three annotators label \need, \asset, and \other~respectively. The text in red shows what can be considered \need~conversation and blue shows \asset~conversation. The final label of \need~was assigned because conversations about \need~forms the majority of the post. This challenging example contextualizes the low agreement between annotators of the CNA dataset.}}
\label{fig:low_alpha_example}
\end{figure}

% leonardelli_agreeing_2021,bugert_breaking_2020,bai_pre-train_2021
\paragraph{Inter-Annotator Agreement}
We measured inter-annotator agreement using the statistic Krippendoff's $\alpha$ to compare agreement between the annotators. The agreement between our three independent annotators per batch was at \textit{0.45}. \icwsm{
The moderate Krippendorff's alpha agreement indicates that identifying \need~and \asset~conversations are inherently subjective, even among human annotators. It also highlights the linguistic challenge of consistently distinguishing such abstract ideas and concepts that can be interpreted differently. There may be an element of subjective perspective or individual bias that shapes how people label these conversations as personal experiences influence perceptions~\cite{hube_understanding_2019}.}

\icwsm{Furthermore, the moderate Krippendorff's alpha also suggests inherent noise in using crowdsourced annotations for this dataset where disagreements may lead to inconsistent labels, especially for borderline cases. As can be seen in the example in Figure~\ref{fig:low_alpha_example}, one comment can have examples of both \asset~and \need~leading to confusion among annotators. The final label for each conversation was decided as the majority label from the three annotators. We further annotated the dataset with one graduate student researcher to break three-way ties in the case of three different labels assigned by the three crowdsourced annotators. The researcher is familiar with the annotation guidelines, and the community needs and assets domain, and has research experience in urban data science. Hence, this annotator was a reliable source of quality control over the annotations by the crowd-workers~\cite{hsueh_data_2009}.}

% The constructs of "needs" and "assets" may be nuanced categories without strictly defined boundaries. There is room for debate on what constitutes a need versus an asset.
% The language used in online informal communication makes categorization difficult. Conversations likely don't neatly fit into academic conceptions of needs and assets.
% Given these sources of ambiguity, we should view classification performance in the 
% context of this fundamental uncertainty, rather than expecting models to achieve high precision. The labels represent imperfect human judgments rather than objective truth. Algorithmic analysis aims to identify linguistic signals within the inherent noise. Future work could study annotator disagreements to reveal insights into concepts of needs and assets across individuals with diverse perspectives.}

\begin{figure}[h]
    \centering
    \includegraphics[width=0.8\linewidth]{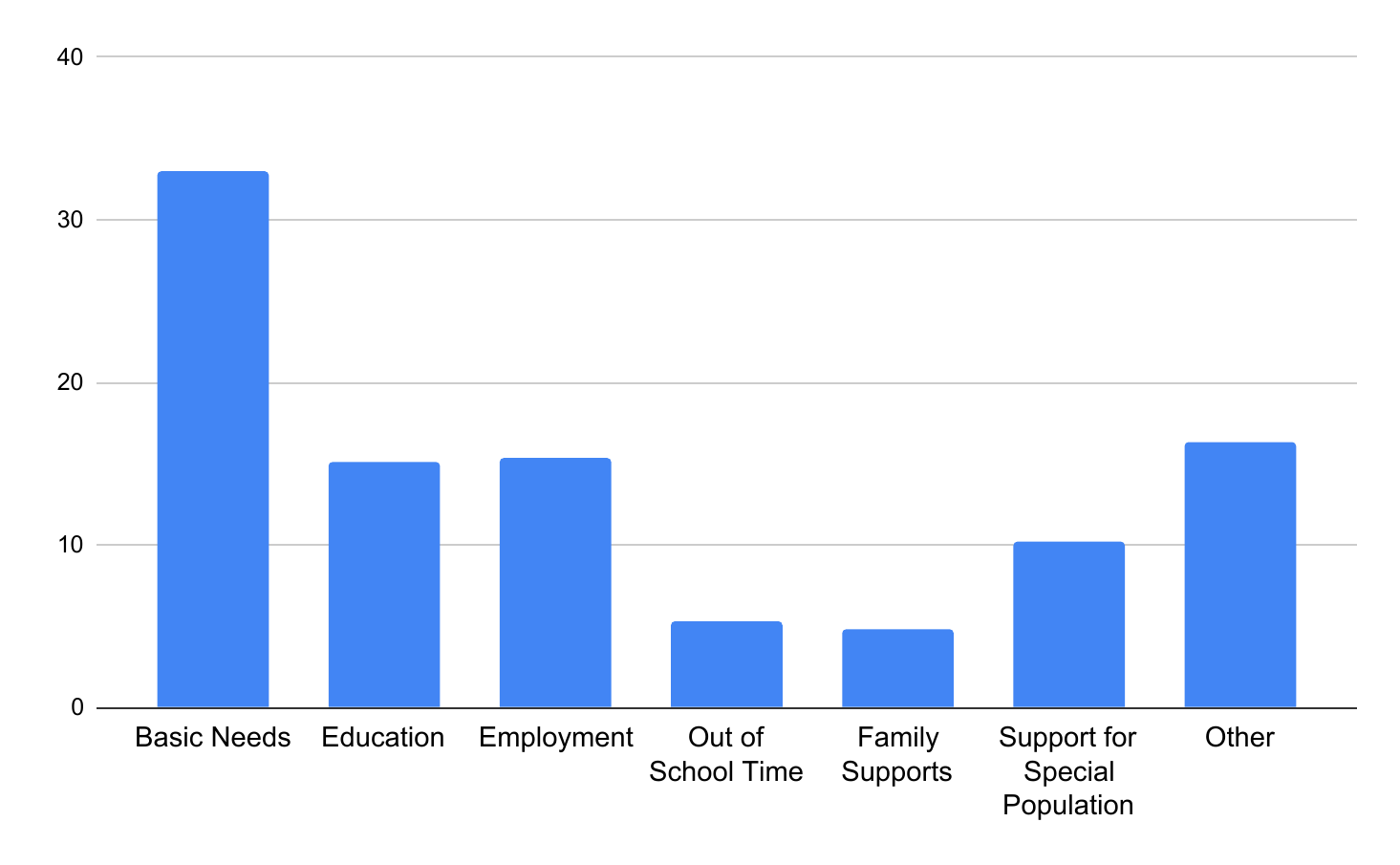}
    \caption{Distribution of categories of needs in conversations in our CNA dataset. The $y$ axis represents the percentage of data categorized as one of the taxonomies.}
    \label{fig:category_needs}
\end{figure}

\paragraph{Categorizing Needs and Assets}
Upon studying the overall distribution of categories within \need~based conversations in Figure~\ref{fig:category_needs}, we see that the annotators defaulted to the categorizing a \need~conversation as \textbf{Basic Needs} in our dataset, indicating that most of the conversations in our dataset are focused on the fundamental necessities in a neighborhood, while the second most popular discussions are between \textbf{Education} and \textbf{Employment}. Upon studying the overall distribution of categories within \asset~conversations in Figure~\ref{fig:categorize_assets}, we see that there is no such default selection for such conversations. There is an even distribution for the type of \asset~within the community conversations in our dataset.

\begin{figure}[h]
    \centering
    \includegraphics[width=0.8\linewidth]{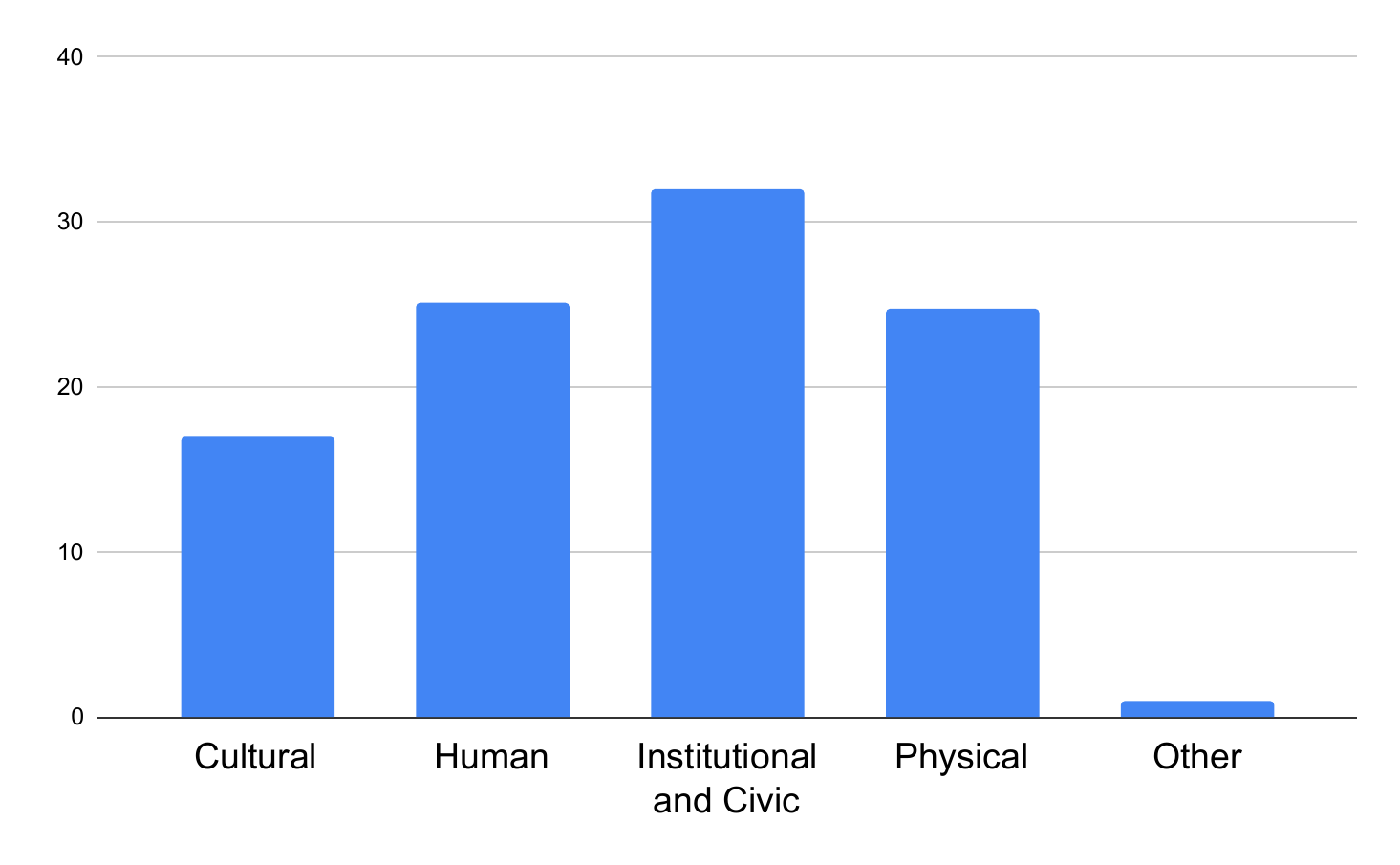}
    \caption{Distribution of categories of assets in conversations in our CNA dataset. The $y$ axis represents the percentage of data categorized as one of the taxonomies.}
    \label{fig:category_assets}
\end{figure}
\section{Classification of Community Needs and Assets}
We define the task of identifying community needs and assets as an utterance-level classification task where a model has to predict
the target labels \need,~\asset, and \other~from each natural language conversation $c_i \in C_{final}$ where $C_{final}$ is the corpus of conversations.

Evaluating the classification of community needs and assets is difficult, as can be seen with the moderate inter-annotator agreement we discovered through our annotation process. 
In order to evaluate the efficacy of our dataset and the separability of needs \& assets as an utterance-level classification task, we designed two baseline approaches to evaluate the classification of community needs and assets from conversations along our utterance-level classification approach.
In this section, we describe each of our baseline methods along with our supervised approach in detail.
 \begin{figure}[t]
     \centering
     \includegraphics[width=0.7\linewidth]{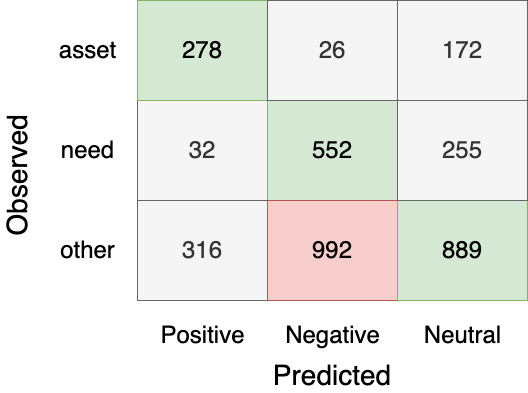}
     \caption{The Confusion Matrix for the Sentiment Classifier. Green indicates True Positive samples, while red is used to highlight negative sentiment in \other~categories.}
     \label{fig:conf_matrix_sentiment}
 \end{figure}

\subsection{Sentiment Classification}
Sentiment classification is a widely used method for opinion mining in online communities~\cite{sv_analysis_2021}. \icwsm{Particularly, sentiment and opinion mining approaches have been used to study particular communities in domains such as health care~\cite{tavoschi_twitter_2020,chintalapudi_text_2021}. Expanding from these approaches, we can assume that} needs-based conversations can be of negative sentiment while asset-based conversations can be of positive sentiment and assign the labels \textcolor{blue}{positive} to comments labeled as \asset,~\icwsm{negative} to comments labeled as \need, and \textcolor{teal}{neutral} to conversations not related to either.
\icwsm{But is that assumption of \need~to negative and \asset~to positive robust enough to perform community needs and asset extraction using baseline sentiment classification? We aim to answer this question using our experiment.}
We use the sentiment classification model derived from a \textit{RoBERTa-base} model trained on approximately 124 million tweets from January 2018 to December 2021 and fine-tuned for sentiment analysis with the TweetEval benchmark~\cite{loureiro_timelms_2022,barbieri_tweeteval_2020}. We utilize a model previously trained on Twitter data to ensure that it can handle noisy social media conversations in our Reddit dataset.

\subsection{LLM Zero-Shot Classification}
We further utilized a zero-shot text classification method using a popular and accessible large language model (LLM) - Google's PaLM2~\cite{chowdhery_palm_2022}. We designed a prompt with the best practices prescribed by~\citet{ziems_can_2023} with a particular focus on mimicking real-life multiple-choice questions. 
Our prompt as shown in Figure~\ref{fig:prompt_class} is the exact same as the question we asked our human annotators. 
Recent studies have shown that LLMs outperform human annotators in labeling tasks \cite{gilardi_chatgpt_2023,huang_is_2023,wang_is_2023}, and our analysis of community needs and assets wanted to accommodate for such findings as well. 
In our usage of PaLM2, we tuned the $temperature$ parameter for text generation to $0.1$ to introduce some non-determinism while also keeping the classification as consistent as possible. We performed a sweep of temperature settings from $0$ to $1$ at an increment of $0.1$ and selected the highest-performing one.

\subsection{Supervised Classification}
We developed a classification model utilizing a pre-trained language model BERT~\cite{devlin_bert_2018}, adding a multi-class text classification layer on top that we fine-tuned using our CNA dataset with target labels \need, \asset, and \other. Furthermore, we also trained classifiers for categories within each of the primary labels. 
We hypothesize that differentiating between \need, \asset, and \other~is possible if the data can be classified to a high degree of accuracy while categorizing further into the taxonomy of needs and assets will be difficult. We trained all our classifiers for $5$ epochs, with an initial learning rate of $5*10^{-5}$ using Adam optimizer~\cite{kingma_adam_2017}, with a training set of 80\% and a validation set of 20\%. We discuss our results in the following section.
\begin{table}[h]
    \centering
    \begin{tabular}{|l r r r r|}
         \hline
         \textbf{method} & \textbf{label} & \textbf{precision} & \textbf{recall} & \textbf{f1-score}  \\
         \hline
         \hline
         Supervised & \asset & 0.93 & 0.94 & 0.94  \\
         {}         &  \need & 0.89 & 0.95 & 0.92  \\
         {}         &  \other & 0.97 & 0.95 & 0.96  \\
         {}         &  macro & 0.93 & 0.95 & 0.94  \\
         {}         &  weighted & 0.95 & 0.95 & 0.95  \\
         \hline
         % \textbf{Sentiment} & {} & {} & {} \\
         {Sentiment} & \asset & 0.44 & 0.58 & 0.50  \\
         {} & \need & 0.35 & 0.66 & 0.46  \\
         {} & \other & 0.68 & 0.40 & 0.51  \\
         {} & macro & 0.49 & 0.55 & 0.49  \\
         {} & weighted & 0.57 & 0.49 & 0.49  \\
         \hline
         % LLM Zero-Shot & {} & {} & {} \\
         {LLM} & \asset & 0.44 & 0.84 & 0.58  \\
         {} & \need & 0.51 & 0.67 & 0.58  \\
         {} & \other & 0.82 & 0.57 & 0.67  \\
         {} & macro & 0.59 & 0.69 & 0.61  \\
         {} & weighted & 0.69 & 0.63 & 0.64  \\
         \hline
    \end{tabular}
    \caption{Results of all the classification model evaluations on the validation set of the CNA dataset. All the results are the mean of 10 runs over the validation set to account for randomness.}
    \label{tab:f1_cna_classification}
\end{table}

\begin{table}[h]
    \centering
    \begin{tabular}{|l rrrr|}
         \hline
         \textbf{class} & \textbf{label} & \textbf{precision} & \textbf{recall} & \textbf{f1-score}  \\
         \hline
         \hline
          {\need} & Basic Needs & 0.82 & 0.92 & 0.87\\
          {} & Education & 0.67 & 0.63 & 0.65\\
          {} & Employment & 0.8 & 0.73 & 0.76\\
          {} & School & 0 & 0 & 0\\
          {} & Family & 0 & 0 & 0\\
          {} & SP & 0.57 & 0.8 & 0.67\\
          {} & Other & 0.66 & 0.75 & 0.7\\
         \hline
         {\asset} & Cultural & 0.86 & 0.5 & 0.63  \\
         {} & Human & 0.67 & 0.92 & 0.78  \\
         {} & IC & 0.84 & 0.84 & 0.84  \\
         {} & Physical & 0.70 & 0.56 & 0.62  \\
         {} & Other & 0.0 & 0.0 & 0.0  \\
         \hline
    \end{tabular}
    \caption{Results for classifying \need~and \asset~categories on the validation set of our CNA dataset. All the results are the mean of 10 runs over the validation set to account for randomness. SP stands for ``Support for Special Population'', IC stands for ``Institutional and Civic'', and School stands for ``Out of School Time''. F1-score of 0 in this table indicates that our model failed to classify the instances.}
    \label{tab:f1_categorization}
\end{table}
\section{Results and Analysis}
\subsection{RQ1: Classifying Needs and Assets Conversations}
\paragraph{Baseline Classification}
The performance of the three classification models is shown in Table~\ref{tab:f1_cna_classification}. Our supervised classification approach outperforms both sentiment and llm zero-shot classification with a macro F1 score of \icwsm{0.94}, while the other two baseline methods struggle at F1 scores of \icwsm{0.49} and \icwsm{0.61} respectively.
% If we look at the performance of the models across the labels, all models perform better in identifying needs-based conversations compared to asset-based conversations. 

\paragraph{Sentiment Analysis}
\icwsm{Our baseline sentiment classification approach in Figure~\ref{fig:conf_matrix_sentiment} shows that 65\% of \need~conversations are classified as \textcolor{red}{negative}, which indicates that \need~conversations have negative sentiment attached to them.
Similarly, 58\% of asset-based conversations have a positive sentiment. To further test the validity of this claim, we perform a chi-squared test of association between (1) \need~conversations and negative sentiment, and (2) \asset~conversation and positive sentiment. The null hypothesis is that there is no association between the two variables while the alternate hypothesis is that there is. We have included the contingency tables for the tests in the appendix. In both cases, the \textit{p-value} is less than 0.01 indicating that we can reject the null hypothesis of no association between the two variables.}

\icwsm{Our classification and hypothesis testing together indicate that while \need~conversations may have negative sentiment, not all negative sentiment conversations are focused on \need. Similarly, \asset~conversations may be statistically inclined towards positive sentiment but not all positive conversations are discussions about community \asset. This simple but important distinction clarifies why in the study of community needs and assets, we need a robust classifier instead of baseline sentiment classification.}
This indicates the difficulty in recognizing assets (compared to needs) and why human annotations are a necessity to build models to identify such conversations about a community.

\paragraph{Categorizing Needs and Assets}
To further deep dive into our labeled \need~and \asset~conversations, we analyzed the performance of our classifier after training it on the further categories of each type of conversation. The results are given in Table~\ref{tab:f1_categorization}. Our \need~category classifiers are capable of detecting Employment and Education labels decently well but it performs best for Basic Needs on our validation set.
Similarly, our \asset~category classifier performs best for Institutional and Civic assets. This phenomenon can be explained due to an existing pattern of identifying entities when talking about assets and highlights in a community.

\subsection{RQ2: Distinguishing between Needs and Assets Conversations}
\paragraph{Emotion Analysis}
\begin{figure}
    \centering
    \includegraphics[width=0.9\linewidth]{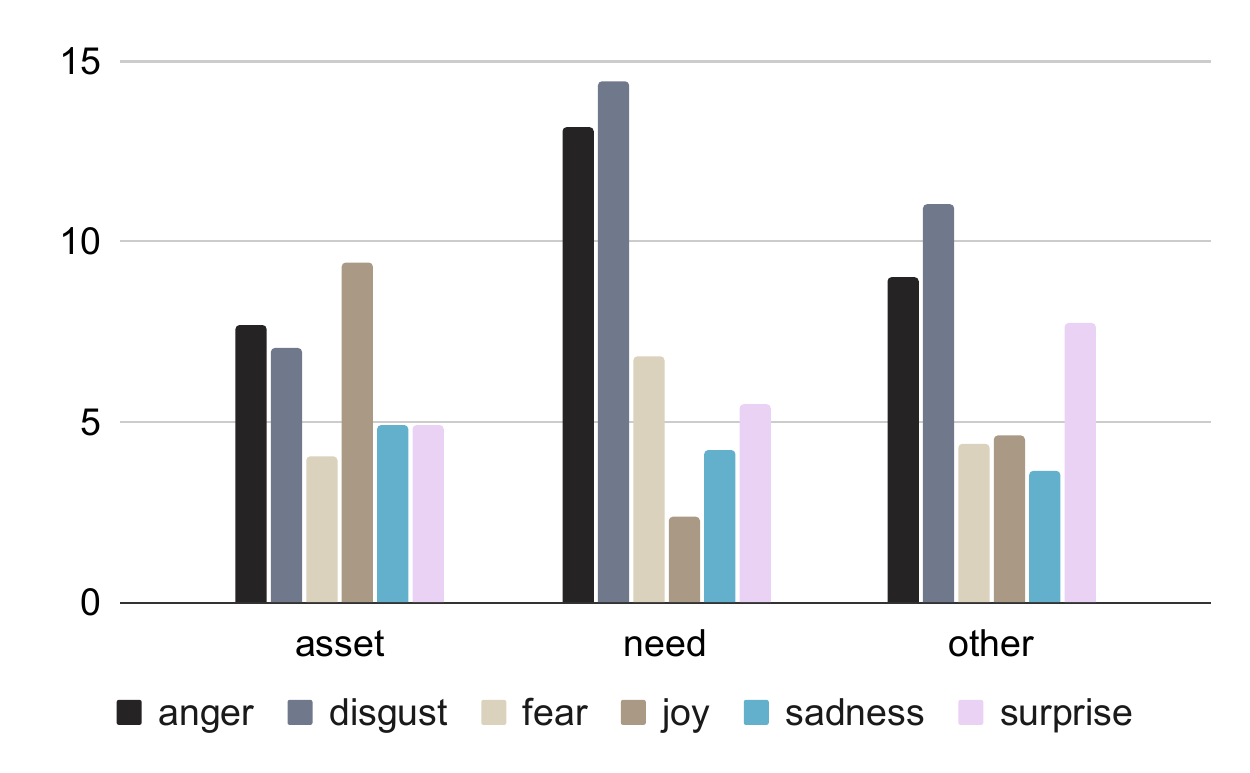}
    \caption{The distribution of emotions for \need, \asset, and \other~classes in the dataset. The $y$ axis displays the percentage of labels that were assigned the emotions. \textit{neutral} emotion was removed as that is the majority at over 50\% for all 3 classes to allow for a cleaner picture.}
    \label{fig:emotional_class}
\end{figure}
We dive further into the sentiment analysis of \need~and \asset~focused conversations by analyzing the emotions of the conversations of each of our labels. We utilize a text classification model fine-tuned on multiple emotion classification datasets~\cite{hartmann_emotion_2022} to classify each of the conversations in our corpus into one of the 6 emotions (and one neutral class). The results shown in Figure~\ref{fig:emotional_class} indicate the distribution of the emotions among our three classes without the neutral class. Neutral is the predominant emotion in all our conversations (\~60\%), indicating that emotion does not play a strong role in the linguistic differences among \need~and \asset~conversations. Removing the neutral class from the Figure~\ref{fig:emotional_class}, it is important to note that \need~conversations have a stronger leaning towards the emotion of \texttt{anger} and \texttt{distrust} and the least amount of \texttt{joy} compared to other labels, while \asset~conversations have a higher percentage of the emotion \texttt{joy} but an equal amount of \texttt{anger} and \texttt{disgust}. 
We have included the figure with the neutral class in our appendix.

\begin{figure}[t]
    \centering
    \includegraphics[width=0.9\linewidth]{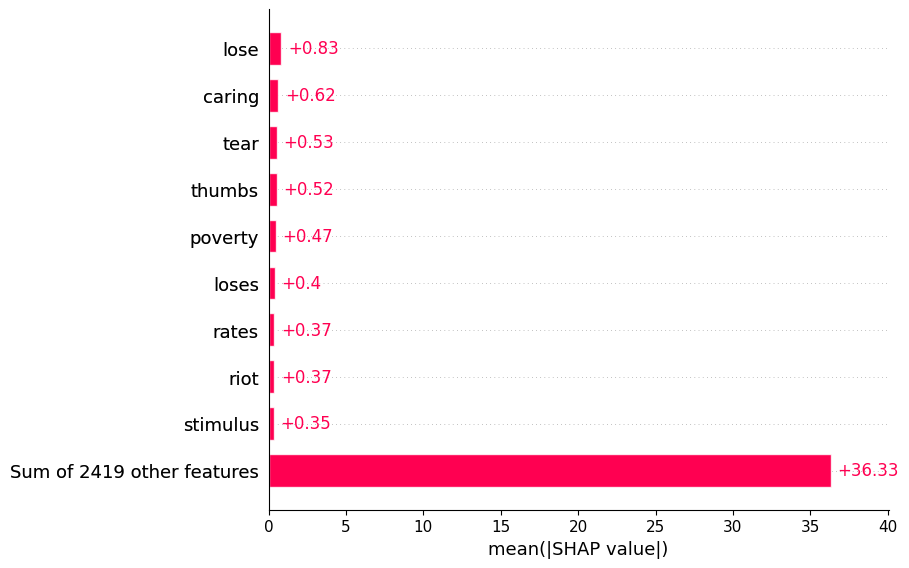}
    \caption{Top 10 word combinations that contribute to the classification of a conversation as \need}
    \label{fig:shap_need}
\end{figure}
\paragraph{Deconstructing Needs and Assets}
We utilized the SHapley Additive exPlanations (SHAP) by~\citet{lundberg_unified_2017} which uses a game-theory inspired method that attempts to enhance the interpretability of machine learning models by computing the importance values for each feature of individual predictions - in our case, the features are words that contribute the most to the classification of each of the labels.

Figure~\ref{fig:shap_need} displays the top 10 words that contribute the most to the classification of 200 samples of conversations classified as \need.
The words with negative connotations such as \textit{poverty}, \textit{lose}, \textit{tear}, appear to have the strongest contributions to the \need~label.

\begin{figure}[t]
    \centering
    \includegraphics[width=0.9\linewidth]{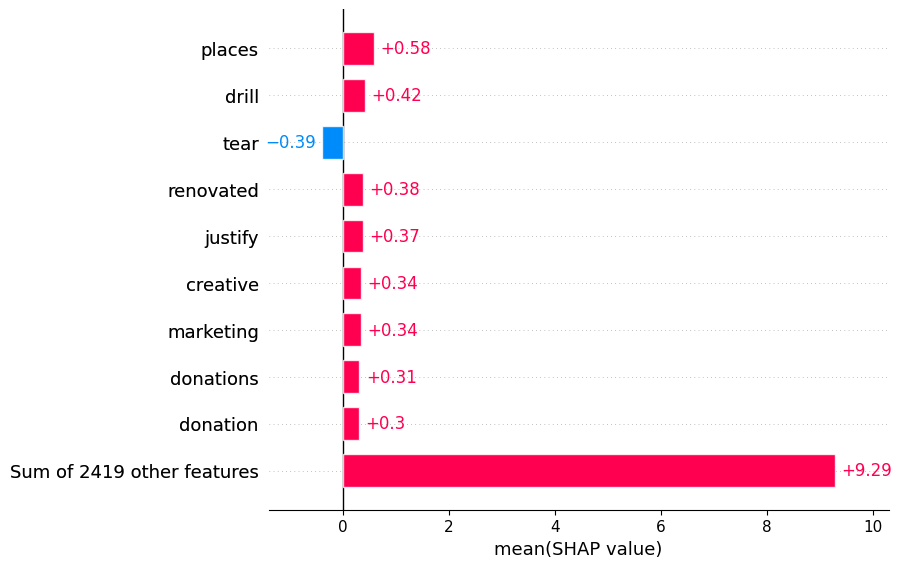}
    \caption{Top 10 word combinations that contribute to the classification of a conversation as \asset}
    \label{fig:shap_asset}
\end{figure}

In contrast, as shown in Figure~\ref{fig:shap_asset} conversations classified as \asset~have a significant focus on words such as \textit{places} and \textit{donations}, words that can relate to identifying a specific entity. This also follows the results we discovered with categorizing \asset~ in Table~\ref{tab:f1_categorization} where we hypothesized that conversations related to Institutional and Civic assets are easier to identify due to the focus on concrete entities. 

\icwsm{We calculate the number of entities in each conversation in the dataset using named entity recognition, particularly limiting entities to type: (1) \texttt{ORG}, (2) \texttt{NORP}, (3) \texttt{GPE}, (4) \texttt{PERSON}, (5) \texttt{WORK\_OF\_ART}. Using the number of entities we perform a one-way ANOVA test to compare the mean number of entities of \asset~conversations relative to all other conversations. Here, the null hypothesis is that the mean number of entities for \asset~conversations is equal to the mean number of entities in all other conversations. The alternate hypothesis is that the mean number of entities is greater for \asset~conversations. At a standard significance level of 5\%, we get the \textit{p-value} at \texttt{0.006} and as a result, we can reject the null hypothesis. This indicates that \asset~conversations may be more location and entity-focused, highlighting the strengths of the community compared to other conversations.}
% \textcolor{red}{TODO: Include a paragraph for hypothesis testing with number of entities. It shows that mean number of entities in asset-based conversations is higher than the mean number of entities in other conversations.}

% An interesting note is that the word \textit{tear}, which contributes positively to the classification of \need, contributes negatively to the classification of \asset.

% \textcolor{red}{TODO: Significance testing missing (R3): There is no significance test for the needs to have negative sentiment. It would have better value to the work to report the significant test. I can see that although the negative is linked to needs, but without having any relation for it nor the positive with other two classes (I mean the assets/others).}

\subsection{Performance In The Wild}
\icwsm{To verify the practicality and robustness of training a classifier using our CNA dataset, we applied our supervised classifier to 23,938 random Reddit conversations from random communities. From these conversations, 1,565 were classified as \asset~conversations and 1,525 were classified as \need~conversations. We randomly picked 50 \asset, 50 \need, and 100 \other~conversations to manually verify the accuracy of our classifier. We measured a macro precision of $0.89$, recall of $0.91$, and f1-score of $0.90$ for in-the-wild performance. The high precision and recall of our classifier demonstrates promising generalization to
unseen data indicating that most classified instances in the annotated sample were true positives.}
\section{Conclusion}
A computational approach to identifying community needs from conversations is an important domain consideration for local non-profits and government agencies to better allocate their resources to foster the development of their communities.
At present, manual approaches are used but with exponentially increasing amounts of data - samples of surveys do not always capture the full picture. To help decrease this manual burden, we take a computational approach to extracting community needs and assets (CNA) and provide a dataset of \icwsm{3,511} related conversations from Reddit with human-annotated labels for \need, \asset, and \other. 
Furthermore, we introduce a baseline community needs and assets (CNA) classification approach and compare it to the zero-shot capabilities of large language models along with sentiment analysis to evaluate the usefulness of our CNA dataset.
Throughout the paper, we answer two research questions:

\paragraph{RQ1: Can we computationally extract community conversations focused on needs and assets?}
Present state-of-the-art sentiment analysis and large language models struggle to differentiate community conversations about \need, \asset, and \other. Classifying and extracting such conversations requires a dataset to fine-tune such machine learning models. Through the use of a supervised approach, we show that it is possible to identify these conversations, and also show that further breakdown and categorization of \need~and \asset~require greater samples of these low-resource conversations.

\paragraph{RQ2: What are the linguistic features that differentiate community needs and assets conversations?}
Our sentiment and emotion analysis on conversations indicated that in our dataset, overall \need~conversations have a slight skew towards negative sentiment along with emotions of \textit{disgust} and \textit{anger}. Similarly, \asset~conversations have the opposite skew towards positive sentiment and include emotions of \textit{joy}. Furthermore, word contributions and hypothesis testing on number of entities in conversations also indicate that asset-based conversations are focused on locations and places indicating that conversations are highlighting a specific entity that can be classified as an \asset.

\paragraph{Future Work}
We plan on expanding our dataset further to incorporate multi-platforms (not just Reddit), and a greater number of conversations such that we can dive even deeper into the categorization of conversations regarding needs and assets. 
Furthermore, we use a simple classification approach; a further study on deep neural networks and their explainability, leveraging the state-of-the-art chain of thought method on zero-shot classification explainability on large language models for example would provide a deeper understanding of the community needs and assets conversation phenomena.
This will increase the robustness of our method and lead the way into an AI-driven approach to community needs assessment that we hope to pilot in the field for exciting results. The future scope of this work can expand not only into real-world applications for non-profits and government agencies, but can also drive theoretical and linguistic understanding of people's wants and needs.
\bibliographystyle{unsrtnat}
\bibliography{references}

\begin{thebibliography}{55}
\providecommand{\natexlab}[1]{#1}
\providecommand{\url}[1]{\texttt{#1}}
\expandafter\ifx\csname urlstyle\endcsname\relax
  \providecommand{\doi}[1]{doi: #1}\else
  \providecommand{\doi}{doi: \begingroup \urlstyle{rm}\Url}\fi

\bibitem[McKnight and Kretzmann(1993)]{mcknight_building_1993}
John McKnight and John Kretzmann.
\newblock Building communities from the inside out.
\newblock \emph{A path toward finding and mobilizing a community’s assets},
  9, 1993.

\bibitem[Witkin and Altschuld(1995)]{witkin_planning_1995}
Belle~Ruth Witkin and James~W Altschuld.
\newblock \emph{Planning and conducting needs assessments: {A} practical
  guide}.
\newblock Sage, 1995.

\bibitem[Billings and Cowley(1995)]{billings_approaches_1995}
Jennifer~Ruth Billings and Sarah Cowley.
\newblock Approaches to community needs assessment: a literature review.
\newblock \emph{Journal of Advanced Nursing}, 22\penalty0 (4):\penalty0
  721--730, 1995.
\newblock ISSN 1365-2648.
\newblock \doi{10.1046/j.1365-2648.1995.22040721.x}.
\newblock URL
  \url{https://onlinelibrary.wiley.com/doi/abs/10.1046/j.1365-2648.1995.22040721.x}.
\newblock \_eprint:
  https://onlinelibrary.wiley.com/doi/pdf/10.1046/j.1365-2648.1995.22040721.x.

\bibitem[Williams et~al.(2020)Williams, Armitage, Tampe, and
  Dienes]{williams_public_2020}
Simon~N Williams, Christopher~J Armitage, Tova Tampe, and Kimberly Dienes.
\newblock Public perceptions and experiences of social distancing and social
  isolation during the {COVID}-19 pandemic: {A} {UK}-based focus group study.
\newblock \emph{BMJ open}, 10\penalty0 (7):\penalty0 e039334, 2020.
\newblock Publisher: British Medical Journal Publishing Group.

\bibitem[Al-Qdah and Lacroix(2017)]{al-qdah_syrian_2017}
Talal Abdel~Karim Al-Qdah and Marie Lacroix.
\newblock Syrian refugees in {Jordan}: {Social} workers use a {Participatory}
  {Rapid} {Appraisal} ({PRA}) methodology for needs assessment, human rights
  and community development.
\newblock \emph{International Social Work}, 60\penalty0 (3):\penalty0 614--627,
  2017.
\newblock Publisher: SAGE Publications Sage UK: London, England.

\bibitem[Alonzo et~al.(2023)Alonzo, Oo, Wijarwadi, and
  Hannigan]{alonzo_using_2023}
Dennis Alonzo, Cherry~Zin Oo, Wendi Wijarwadi, and Caitlin Hannigan.
\newblock Using social media for assessment purposes: {Practices} and future
  directions.
\newblock \emph{Frontiers in Psychology}, 13, 2023.
\newblock ISSN 1664-1078.
\newblock URL
  \url{https://www.frontiersin.org/articles/10.3389/fpsyg.2022.1075818}.

\bibitem[White et~al.(2023)White, Burns, Carson, and Scott]{white_mapping_2023}
Becky~K White, Sharyn~K Burns, Jennie Carson, and Jane~A Scott.
\newblock Mapping breastfeeding and {COVID}-19 related content and engagement
  on {Facebook}: {Results} from an online social listening study.
\newblock \emph{Health Promotion Journal of Australia}, 2023.
\newblock Publisher: Wiley Online Library.

\bibitem[Iqbal et~al.(2023)Iqbal, Ghafouri, Tyson, Suarez-Tangil, and
  Castro]{iqbal_lady_2023}
Waleed Iqbal, Vahid Ghafouri, Gareth Tyson, Guillermo Suarez-Tangil, and
  Ignacio Castro.
\newblock Lady and the {Tramp} {Nextdoor}: {Online} {Manifestations} of
  {Real}-{World} {Inequalities} in the {Nextdoor} {Social} {Network}.
\newblock \emph{Proceedings of the International AAAI Conference on Web and
  Social Media}, 17:\penalty0 399--410, June 2023.
\newblock ISSN 2334-0770.
\newblock \doi{10.1609/icwsm.v17i1.22155}.
\newblock URL \url{https://ojs.aaai.org/index.php/ICWSM/article/view/22155}.

\bibitem[Lasri et~al.(2023)Lasri, Tonneau, Naushan, Malhotra, Farouq,
  Orozco-Olvera, and Fraiberger]{lasri_large-scale_2023}
Karim Lasri, Manuel Tonneau, Haaya Naushan, Niyati Malhotra, Ibrahim Farouq,
  Víctor Orozco-Olvera, and Samuel Fraiberger.
\newblock Large-{Scale} {Demographic} {Inference} of {Social} {Media} {Users}
  in a {Low}-{Resource} {Scenario}.
\newblock \emph{Proceedings of the International AAAI Conference on Web and
  Social Media}, 17:\penalty0 519--529, June 2023.
\newblock ISSN 2334-0770.
\newblock \doi{10.1609/icwsm.v17i1.22165}.
\newblock URL \url{https://ojs.aaai.org/index.php/ICWSM/article/view/22165}.

\bibitem[Zohrabi(2013)]{zohrabi_mixed_2013}
Mohammad Zohrabi.
\newblock Mixed method research: {Instruments}, validity, reliability and
  reporting findings.
\newblock \emph{Theory and practice in language studies}, 3\penalty0
  (2):\penalty0 254, 2013.
\newblock Publisher: Academy Publication Co., Ltd.

\bibitem[Shah et~al.(2015)Shah, Cappella, and Neuman]{shah_big_2015}
Dhavan~V. Shah, Joseph~N. Cappella, and W.~Russell Neuman.
\newblock Big {Data}, {Digital} {Media}, and {Computational} {Social}
  {Science}: {Possibilities} and {Perils}.
\newblock \emph{The Annals of the American Academy of Political and Social
  Science}, 659:\penalty0 6--13, 2015.
\newblock ISSN 0002-7162.
\newblock URL \url{https://www.jstor.org/stable/24541845}.
\newblock Publisher: [Sage Publications, Inc., American Academy of Political
  and Social Science].

\bibitem[Ziems et~al.(2023)Ziems, Held, Shaikh, Chen, Zhang, and
  Yang]{ziems_can_2023}
Caleb Ziems, William Held, Omar Shaikh, Jiaao Chen, Zhehao Zhang, and Diyi
  Yang.
\newblock Can {Large} {Language} {Models} {Transform} {Computational} {Social}
  {Science}?, April 2023.
\newblock URL \url{http://arxiv.org/abs/2305.03514}.
\newblock arXiv:2305.03514 [cs].

\bibitem[S.V. and Ittamalla(2021)]{sv_analysis_2021}
Praveen S.V. and Rajesh Ittamalla.
\newblock An analysis of attitude of general public toward {COVID}-19 crises
  – sentimental analysis and a topic modeling study.
\newblock \emph{Information Discovery and Delivery}, 49\penalty0 (3):\penalty0
  240--249, January 2021.
\newblock ISSN 2398-6247.
\newblock \doi{10.1108/IDD-08-2020-0097}.
\newblock URL \url{https://doi.org/10.1108/IDD-08-2020-0097}.
\newblock Publisher: Emerald Publishing Limited.

\bibitem[Mejova and Manikonda(2023)]{mejova_comfort_2023}
Yelena Mejova and Lydia Manikonda.
\newblock Comfort {Foods} and {Community} {Connectedness}: {Investigating}
  {Diet} {Change} during {COVID}-19 {Using} {YouTube} {Videos} on {Twitter}.
\newblock \emph{Proceedings of the International AAAI Conference on Web and
  Social Media}, 17:\penalty0 602--613, June 2023.
\newblock ISSN 2334-0770.
\newblock \doi{10.1609/icwsm.v17i1.22172}.
\newblock URL \url{https://ojs.aaai.org/index.php/ICWSM/article/view/22172}.

\bibitem[Bakhtin(2010)]{bakhtin_speech_2010}
M.~M. Bakhtin.
\newblock \emph{Speech {Genres} and {Other} {Late} {Essays}}.
\newblock University of Texas Press, March 2010.
\newblock ISBN 978-0-292-78287-7.
\newblock Google-Books-ID: n7xaBAAAQBAJ.

\bibitem[Kingsley et~al.(2014)Kingsley, Coulton, and
  Pettit]{kingsley_strengthening_2014}
G~Thomas Kingsley, Claudia~J Coulton, and Kathryn~LS Pettit.
\newblock \emph{Strengthening communities with neighborhood data}.
\newblock Urban Institute Washington, DC, 2014.

\bibitem[Chowdhury and Sharma(2021)]{chowdhury_citizenly_2021}
Md~Towhidul~Absar Chowdhury and Naveen Sharma.
\newblock Citizenly: {A} platform to encourage data-driven decision making for
  the community by the community.
\newblock \emph{2021 IEEE International Conferences on Internet of Things
  (iThings) and IEEE Green Computing \& Communications (GreenCom) and IEEE
  Cyber, Physical \& Social Computing (CPSCom) and IEEE Smart Data (SmartData)
  and IEEE Congress on Cybermatics (Cybermatics)}, pages 359--364, 2021.
\newblock Publisher: IEEE.

\bibitem[Vinuesa et~al.(2020)Vinuesa, Azizpour, Leite, Balaam, Dignum, Domisch,
  Felländer, Langhans, Tegmark, and Fuso~Nerini]{vinuesa_role_2020}
Ricardo Vinuesa, Hossein Azizpour, Iolanda Leite, Madeline Balaam, Virginia
  Dignum, Sami Domisch, Anna Felländer, Simone~Daniela Langhans, Max Tegmark,
  and Francesco Fuso~Nerini.
\newblock The role of artificial intelligence in achieving the {Sustainable}
  {Development} {Goals}.
\newblock \emph{Nature Communications}, 11\penalty0 (1):\penalty0 233, January
  2020.
\newblock ISSN 2041-1723.
\newblock \doi{10.1038/s41467-019-14108-y}.
\newblock URL \url{https://www.nature.com/articles/s41467-019-14108-y}.
\newblock Number: 1 Publisher: Nature Publishing Group.

\bibitem[Althoff et~al.(2014)Althoff, Danescu-Niculescu-Mizil, and
  Jurafsky]{althoff_how_2014}
Tim Althoff, Cristian Danescu-Niculescu-Mizil, and Dan Jurafsky.
\newblock How to {Ask} for a {Favor}: {A} {Case} {Study} on the {Success} of
  {Altruistic} {Requests}.
\newblock \emph{Proceedings of the International AAAI Conference on Web and
  Social Media}, 8\penalty0 (1):\penalty0 12--21, May 2014.
\newblock ISSN 2334-0770.
\newblock \doi{10.1609/icwsm.v8i1.14547}.
\newblock URL \url{https://ojs.aaai.org/index.php/ICWSM/article/view/14547}.
\newblock Number: 1.

\bibitem[Cui et~al.(2022)Cui, Zhang, Jaidka, Pang, Sherman, Jakhetiya, Ungar,
  and Guntuku]{cui_social_2022}
Jesse Cui, Tingdan Zhang, Kokil Jaidka, Dandan Pang, Garrick Sherman, Vinit
  Jakhetiya, Lyle~H. Ungar, and Sharath~Chandra Guntuku.
\newblock Social {Media} {Reveals} {Urban}-{Rural} {Differences} in {Stress}
  across {China}.
\newblock \emph{Proceedings of the International AAAI Conference on Web and
  Social Media}, 16:\penalty0 114--124, May 2022.
\newblock ISSN 2334-0770.
\newblock \doi{10.1609/icwsm.v16i1.19277}.
\newblock URL \url{https://ojs.aaai.org/index.php/ICWSM/article/view/19277}.

\bibitem[Giorgi et~al.(2023)Giorgi, Zhao, Feng, and Martin]{giorgi_author_2023}
Salvatore Giorgi, Ke~Zhao, Alexander~H. Feng, and Lara~J. Martin.
\newblock Author as {Character} and {Narrator}: {Deconstructing} {Personal}
  {Narratives} from the r/{AmITheAsshole} {Reddit} {Community}.
\newblock \emph{Proceedings of the International AAAI Conference on Web and
  Social Media}, 17:\penalty0 233--244, June 2023.
\newblock ISSN 2334-0770.
\newblock \doi{10.1609/icwsm.v17i1.22141}.
\newblock URL \url{https://ojs.aaai.org/index.php/ICWSM/article/view/22141}.

\bibitem[Herdağdelen et~al.(2023)Herdağdelen, Adamic, and
  State]{herdagdelen_geography_2023}
Amaç Herdağdelen, Lada Adamic, and Bogdan State.
\newblock The {Geography} of {Facebook} {Groups} in the {United} {States}.
\newblock \emph{Proceedings of the International AAAI Conference on Web and
  Social Media}, 17:\penalty0 351--362, June 2023.
\newblock ISSN 2334-0770.
\newblock \doi{10.1609/icwsm.v17i1.22151}.
\newblock URL \url{https://ojs.aaai.org/index.php/ICWSM/article/view/22151}.

\bibitem[Demszky et~al.(2019)Demszky, Garg, Voigt, Zou, Shapiro, Gentzkow, and
  Jurafsky]{demszky_analyzing_2019}
Dorottya Demszky, Nikhil Garg, Rob Voigt, James Zou, Jesse Shapiro, Matthew
  Gentzkow, and Dan Jurafsky.
\newblock Analyzing {Polarization} in {Social} {Media}: {Method} and
  {Application} to {Tweets} on 21 {Mass} {Shootings}.
\newblock In \emph{Proceedings of the 2019 {Conference} of the {North}
  {American} {Chapter} of the {Association} for {Computational} {Linguistics}:
  {Human} {Language} {Technologies}, {Volume} 1 ({Long} and {Short} {Papers})},
  pages 2970--3005, Minneapolis, Minnesota, June 2019. Association for
  Computational Linguistics.
\newblock \doi{10.18653/v1/N19-1304}.
\newblock URL \url{https://aclanthology.org/N19-1304}.

\bibitem[Ortony et~al.(2022)Ortony, Clore, and Collins]{ortony_cognitive_2022}
Andrew Ortony, Gerald~L Clore, and Allan Collins.
\newblock \emph{The cognitive structure of emotions}.
\newblock Cambridge university press, 2022.

\bibitem[ElSherief et~al.(2021)ElSherief, Ziems, Muchlinski, Anupindi, Seybolt,
  De~Choudhury, and Yang]{elsherief_latent_2021}
Mai ElSherief, Caleb Ziems, David Muchlinski, Vaishnavi Anupindi, Jordyn
  Seybolt, Munmun De~Choudhury, and Diyi Yang.
\newblock Latent {Hatred}: {A} {Benchmark} for {Understanding} {Implicit}
  {Hate} {Speech}, September 2021.
\newblock URL \url{http://arxiv.org/abs/2109.05322}.
\newblock arXiv:2109.05322 [cs].

\bibitem[Dutta et~al.(2022)Dutta, Li, Nagin, and
  KhudaBukhsh]{dutta_murder_2022}
Sujan Dutta, Beibei Li, Daniel~S. Nagin, and Ashiqur~R. KhudaBukhsh.
\newblock A {Murder} and {Protests}, the {Capitol} {Riot}, and the {Chauvin}
  {Trial}: {Estimating} {Disparate} {News} {Media} {Stance}.
\newblock In \emph{Proceedings of the {Thirty}-{First} {IJCAI}}, pages
  5059--5065, Vienna, Austria, July 2022. International Joint Conferences on
  Artificial Intelligence Organization.
\newblock ISBN 978-1-956792-00-3.
\newblock \doi{10.24963/ijcai.2022/702}.
\newblock URL \url{https://www.ijcai.org/proceedings/2022/702}.

\bibitem[Alam et~al.(2021)Alam, Shaar, Dalvi, Sajjad, Nikolov, Mubarak,
  Da~San~Martino, Abdelali, Durrani, Darwish, Al-Homaid, Zaghouani, Caselli,
  Danoe, Stolk, Bruntink, and Nakov]{alam_fighting_2021}
Firoj Alam, Shaden Shaar, Fahim Dalvi, Hassan Sajjad, Alex Nikolov, Hamdy
  Mubarak, Giovanni Da~San~Martino, Ahmed Abdelali, Nadir Durrani, Kareem
  Darwish, Abdulaziz Al-Homaid, Wajdi Zaghouani, Tommaso Caselli, Gijs Danoe,
  Friso Stolk, Britt Bruntink, and Preslav Nakov.
\newblock Fighting the {COVID}-19 {Infodemic}: {Modeling} the {Perspective} of
  {Journalists}, {Fact}-{Checkers}, {Social} {Media} {Platforms}, {Policy}
  {Makers}, and the {Society}.
\newblock In \emph{Findings of the {Association} for {Computational}
  {Linguistics}: {EMNLP} 2021}, pages 611--649, Punta Cana, Dominican Republic,
  November 2021. Association for Computational Linguistics.
\newblock \doi{10.18653/v1/2021.findings-emnlp.56}.
\newblock URL \url{https://aclanthology.org/2021.findings-emnlp.56}.

\bibitem[Jiang et~al.(2023)Jiang, Ren, and Ferrara]{jiang_retweet-bert_2023}
Julie Jiang, Xiang Ren, and Emilio Ferrara.
\newblock Retweet-{BERT}: {Political} {Leaning} {Detection} {Using} {Language}
  {Features} and {Information} {Diffusion} on {Social} {Networks}.
\newblock \emph{Proceedings of the International AAAI Conference on Web and
  Social Media}, 17:\penalty0 459--469, June 2023.
\newblock ISSN 2334-0770.
\newblock \doi{10.1609/icwsm.v17i1.22160}.
\newblock URL \url{https://ojs.aaai.org/index.php/ICWSM/article/view/22160}.

\bibitem[DYCD(2022)]{nyc_dycd_community_2022}
NYC DYCD.
\newblock Community {Needs} {Assessment} ({CNA}) - {DYCD}, 2022.
\newblock URL
  \url{https://www.nyc.gov/site/dycd/involved/boards-and-councils/CNA.page}.

\bibitem[Flora et~al.(2016)Flora, Flora, and Gasteyer]{flora_rural_2016}
Cornelia~Butler Flora, Jan~L. Flora, and Stephen~P. Gasteyer.
\newblock \emph{Rural {Communities}: {Legacy} + {Change}}.
\newblock Avalon Publishing, 2016.
\newblock ISBN 978-0-8133-4971-8.
\newblock Google-Books-ID: BMQVCgAAQBAJ.

\bibitem[Chowdhury and Sharma(2022)]{chowdhury_community_2022}
Md~Towhidul~Absar Chowdhury and Naveen Sharma.
\newblock Community {Asset} {Ontology} for {Modeling} {Community} {Data} using
  {Information} {Extraction}.
\newblock In \emph{2022 6th {International} {Conference} on {Natural}
  {Language} {Processing} and {Information} {Retrieval} ({NLPIR})}, {NLPIR}
  2022, pages 1--5, New York, NY, USA, 2022. Association for Computing
  Machinery.
\newblock event-place: Bangkok, Thailand.

\bibitem[Callaghan and Colton(2008)]{callaghan_building_2008}
Edith~G Callaghan and John Colton.
\newblock Building sustainable \& resilient communities: a balancing of
  community capital.
\newblock \emph{Environment, development and sustainability}, 10\penalty0
  (6):\penalty0 931--942, 2008.
\newblock Publisher: Springer.

\bibitem[Schultz et~al.(2000)Schultz, Fawcett, Francisco, Wolff, Berkowitz, and
  Nagy]{schultz_community_2000}
Jerry~A Schultz, Stephen~B Fawcett, Vincent~T Francisco, Tom Wolff, Bill~R
  Berkowitz, and Genevieve Nagy.
\newblock The {Community} {Tool} {Box}: {Using} the {Internet} to support the
  work of community health and development.
\newblock \emph{Journal of Technology in Human Services}, 17\penalty0
  (2-3):\penalty0 193--215, 2000.
\newblock Publisher: Taylor \& Francis.

\bibitem[Lokala et~al.(2022)Lokala, Srivastava, Dastidar, Chakraborty, Akhtar,
  Panahiazar, and Sheth]{lokala_computational_2022}
Usha Lokala, Aseem Srivastava, Triyasha~Ghosh Dastidar, Tanmoy Chakraborty,
  Md~Shad Akhtar, Maryam Panahiazar, and Amit Sheth.
\newblock A {Computational} {Approach} to {Understand} {Mental} {Health} from
  {Reddit}: {Knowledge}-{Aware} {Multitask} {Learning} {Framework}.
\newblock \emph{Proceedings of the International AAAI Conference on Web and
  Social Media}, 16:\penalty0 640--650, May 2022.
\newblock ISSN 2334-0770.
\newblock \doi{10.1609/icwsm.v16i1.19322}.
\newblock URL \url{https://ojs.aaai.org/index.php/ICWSM/article/view/19322}.

\bibitem[Tausanovitch and Warshaw(2014)]{tausanovitch_representation_2014}
Chris Tausanovitch and Christopher Warshaw.
\newblock Representation in municipal government.
\newblock \emph{American Political Science Review}, 108\penalty0 (3):\penalty0
  605--641, 2014.
\newblock Publisher: Cambridge University Press.

\bibitem[Halterman et~al.(2021)Halterman, Keith, Sarwar, and
  O'Connor]{halterman_corpus-level_2021}
Andrew Halterman, Katherine Keith, Sheikh Sarwar, and Brendan O'Connor.
\newblock Corpus-{Level} {Evaluation} for {Event} {QA}: {The}
  {IndiaPoliceEvents} {Corpus} {Covering} the 2002 {Gujarat} {Violence}.
\newblock In \emph{Findings of the {Association} for {Computational}
  {Linguistics}: {ACL}-{IJCNLP} 2021}, pages 4240--4253, Online, August 2021.
  Association for Computational Linguistics.
\newblock \doi{10.18653/v1/2021.findings-acl.371}.
\newblock URL \url{https://aclanthology.org/2021.findings-acl.371}.

\bibitem[Khudabukhsh et~al.(2022)Khudabukhsh, Sarkar, Kamlet, and
  Mitchell]{khudabukhsh_fringe_2022}
Ashiqur~R. Khudabukhsh, Rupak Sarkar, Mark~S. Kamlet, and Tom~M. Mitchell.
\newblock Fringe {News} {Networks}: {Dynamics} of {US} {News} {Viewership}
  following the 2020 {Presidential} {Election}.
\newblock In \emph{Proceedings of the 14th {ACM} {Web} {Science} {Conference}
  2022}, {WebSci} '22, pages 269--278, New York, NY, USA, June 2022.
  Association for Computing Machinery.
\newblock ISBN 978-1-4503-9191-7.
\newblock \doi{10.1145/3501247.3531577}.
\newblock URL \url{https://dl.acm.org/doi/10.1145/3501247.3531577}.

\bibitem[Chowdhury et~al.(2024)Chowdhury, Datta, Sharma, and
  KhudaBukhsh]{chowdhury2024infrastructure}
Md~Towhidul~Absar Chowdhury, Soumyajit Datta, Naveen Sharma, and Ashiqur~R
  KhudaBukhsh.
\newblock Infrastructure ombudsman: {Mining} future failure concerns from
  structural disaster response.
\newblock \emph{arXiv preprint arXiv:2402.13528}, 2024.

\bibitem[Leonardelli et~al.(2021)Leonardelli, Menini, Aprosio, Guerini, and
  Tonelli]{leonardelli_agreeing_2021}
Elisa Leonardelli, Stefano Menini, Alessio~Palmero Aprosio, Marco Guerini, and
  Sara Tonelli.
\newblock Agreeing to {Disagree}: {Annotating} {Offensive} {Language}
  {Datasets} with {Annotators}’ {Disagreement}.
\newblock \emph{ArXiv}, abs/2109.13563, 2021.

\bibitem[Bugert et~al.(2020)Bugert, Reimers, Barhom, Dagan, and
  Gurevych]{bugert_breaking_2020}
Michael Bugert, Nils Reimers, Shany Barhom, Ido Dagan, and Iryna Gurevych.
\newblock Breaking the {Subtopic} {Barrier} in {Cross}-{Document} {Event}
  {Coreference} {Resolution}.
\newblock In \emph{{Text2Story}@{ECIR}}, 2020.

\bibitem[Bai et~al.(2021)Bai, Ritter, and Xu]{bai_pre-train_2021}
Fan Bai, Alan Ritter, and Wei Xu.
\newblock Pre-train or {Annotate}? {Domain} {Adaptation} with a {Constrained}
  {Budget}.
\newblock \emph{Proceedings of the 2021 EMNLP}, pages 5002--5015, November
  2021.
\newblock \doi{10.18653/v1/2021.emnlp-main.409}.
\newblock URL \url{https://aclanthology.org/2021.emnlp-main.409}.
\newblock Place: Online and Punta Cana, Dominican Republic Publisher:
  Association for Computational Linguistics.

\bibitem[Hube et~al.(2019)Hube, Fetahu, and Gadiraju]{hube_understanding_2019}
Christoph Hube, Besnik Fetahu, and Ujwal Gadiraju.
\newblock Understanding and {Mitigating} {Worker} {Biases} in the
  {Crowdsourced} {Collection} of {Subjective} {Judgments}.
\newblock In \emph{Proceedings of the 2019 {CHI} {Conference} on {Human}
  {Factors} in {Computing} {Systems}}, {CHI} '19, pages 1--12, New York, NY,
  USA, May 2019. Association for Computing Machinery.
\newblock ISBN 978-1-4503-5970-2.
\newblock \doi{10.1145/3290605.3300637}.
\newblock URL \url{https://dl.acm.org/doi/10.1145/3290605.3300637}.

\bibitem[Hsueh et~al.(2009)Hsueh, Melville, and Sindhwani]{hsueh_data_2009}
Pei-Yun Hsueh, Prem Melville, and Vikas Sindhwani.
\newblock Data quality from crowdsourcing: a study of annotation selection
  criteria.
\newblock In \emph{Proceedings of the {NAACL} {HLT} 2009 workshop on active
  learning for natural language processing}, pages 27--35, 2009.

\bibitem[Tavoschi et~al.(2020)Tavoschi, Quattrone, D’Andrea, Ducange,
  Vabanesi, Marcelloni, and Lopalco]{tavoschi_twitter_2020}
Lara Tavoschi, Filippo Quattrone, Eleonora D’Andrea, Pietro Ducange, Marco
  Vabanesi, Francesco Marcelloni, and Pier~Luigi Lopalco.
\newblock Twitter as a sentinel tool to monitor public opinion on vaccination:
  an opinion mining analysis from {September} 2016 to {August} 2017 in {Italy}.
\newblock \emph{Human Vaccines \& Immunotherapeutics}, 16\penalty0
  (5):\penalty0 1062--1069, May 2020.
\newblock ISSN 2164-5515.
\newblock \doi{10.1080/21645515.2020.1714311}.
\newblock URL \url{https://doi.org/10.1080/21645515.2020.1714311}.
\newblock Publisher: Taylor \& Francis \_eprint:
  https://doi.org/10.1080/21645515.2020.1714311.

\bibitem[Chintalapudi et~al.(2021)Chintalapudi, Battineni, Canio, Sagaro, and
  Amenta]{chintalapudi_text_2021}
Nalini Chintalapudi, Gopi Battineni, Marzio~Di Canio, Getu~Gamo Sagaro, and
  Francesco Amenta.
\newblock Text mining with sentiment analysis on seafarers’ medical
  documents.
\newblock \emph{International Journal of Information Management Data Insights},
  1\penalty0 (1):\penalty0 100005, April 2021.
\newblock ISSN 2667-0968.
\newblock \doi{10.1016/j.jjimei.2020.100005}.
\newblock URL
  \url{https://www.sciencedirect.com/science/article/pii/S2667096820300057}.

\bibitem[Loureiro et~al.(2022)Loureiro, Barbieri, Neves, Espinosa~Anke, and
  Camacho-collados]{loureiro_timelms_2022}
Daniel Loureiro, Francesco Barbieri, Leonardo Neves, Luis Espinosa~Anke, and
  Jose Camacho-collados.
\newblock {TimeLMs}: {Diachronic} {Language} {Models} from {Twitter}.
\newblock In \emph{Proceedings of the 60th {Annual} {Meeting} of the
  {Association} for {Computational} {Linguistics}: {System} {Demonstrations}},
  pages 251--260, Dublin, Ireland, May 2022. Association for Computational
  Linguistics.
\newblock \doi{10.18653/v1/2022.acl-demo.25}.
\newblock URL \url{https://aclanthology.org/2022.acl-demo.25}.

\bibitem[Barbieri et~al.(2020)Barbieri, Camacho-Collados, Espinosa~Anke, and
  Neves]{barbieri_tweeteval_2020}
Francesco Barbieri, Jose Camacho-Collados, Luis Espinosa~Anke, and Leonardo
  Neves.
\newblock {TweetEval}: {Unified} {Benchmark} and {Comparative} {Evaluation} for
  {Tweet} {Classification}.
\newblock In \emph{Findings of the {Association} for {Computational}
  {Linguistics}: {EMNLP} 2020}, pages 1644--1650, Online, November 2020.
  Association for Computational Linguistics.
\newblock \doi{10.18653/v1/2020.findings-emnlp.148}.
\newblock URL \url{https://aclanthology.org/2020.findings-emnlp.148}.

\bibitem[Chowdhery et~al.(2022)Chowdhery, Narang, Devlin, Bosma, Mishra,
  Roberts, Barham, Chung, Sutton, Gehrmann, and {others}]{chowdhery_palm_2022}
Aakanksha Chowdhery, Sharan Narang, Jacob Devlin, Maarten Bosma, Gaurav Mishra,
  Adam Roberts, Paul Barham, Hyung~Won Chung, Charles Sutton, Sebastian
  Gehrmann, and {others}.
\newblock Palm: {Scaling} language modeling with pathways.
\newblock \emph{arXiv preprint arXiv:2204.02311}, 2022.

\bibitem[Gilardi et~al.(2023)Gilardi, Alizadeh, and
  Kubli]{gilardi_chatgpt_2023}
Fabrizio Gilardi, Meysam Alizadeh, and Maël Kubli.
\newblock {ChatGPT} {Outperforms} {Crowd}-{Workers} for {Text}-{Annotation}
  {Tasks}, March 2023.
\newblock URL \url{http://arxiv.org/abs/2303.15056}.
\newblock arXiv:2303.15056 [cs].

\bibitem[Huang et~al.(2023)Huang, Kwak, and An]{huang_is_2023}
Fan Huang, Haewoon Kwak, and Jisun An.
\newblock Is {ChatGPT} better than {Human} {Annotators}? {Potential} and
  {Limitations} of {ChatGPT} in {Explaining} {Implicit} {Hate} {Speech}.
\newblock In \emph{Companion {Proceedings} of the {ACM} {Web} {Conference}
  2023}, pages 294--297, April 2023.
\newblock \doi{10.1145/3543873.3587368}.
\newblock URL \url{http://arxiv.org/abs/2302.07736}.
\newblock arXiv:2302.07736 [cs].

\bibitem[Wang et~al.(2023)Wang, Liang, Meng, Sun, Shi, Li, Xu, Qu, and
  Zhou]{wang_is_2023}
Jiaan Wang, Yunlong Liang, Fandong Meng, Zengkui Sun, Haoxiang Shi, Zhixu Li,
  Jinan Xu, Jianfeng Qu, and Jie Zhou.
\newblock Is {ChatGPT} a {Good} {NLG} {Evaluator}? {A} {Preliminary} {Study},
  April 2023.
\newblock URL \url{http://arxiv.org/abs/2303.04048}.
\newblock arXiv:2303.04048 [cs].

\bibitem[Devlin et~al.(2018)Devlin, Chang, Lee, and
  Toutanova]{devlin_bert_2018}
Jacob Devlin, Ming-Wei Chang, Kenton Lee, and Kristina Toutanova.
\newblock Bert: {Pre}-training of deep bidirectional transformers for language
  understanding.
\newblock \emph{arXiv preprint arXiv:1810.04805}, 2018.

\bibitem[Kingma and Ba(2017)]{kingma_adam_2017}
Diederik~P. Kingma and Jimmy Ba.
\newblock Adam: {A} {Method} for {Stochastic} {Optimization}, January 2017.
\newblock URL \url{http://arxiv.org/abs/1412.6980}.
\newblock arXiv:1412.6980 [cs].

\bibitem[Hartmann(2022)]{hartmann_emotion_2022}
Jochen Hartmann.
\newblock Emotion {English} {DistilRoBERTa}-base, 2022.
\newblock URL
  \url{https://huggingface.co/j-hartmann/emotion-english-distilroberta-base/}.

\bibitem[Lundberg and Lee(2017)]{lundberg_unified_2017}
Scott~M. Lundberg and Su-In Lee.
\newblock A unified approach to interpreting model predictions.
\newblock In \emph{Proceedings of the 31st {International} {Conference} on
  {Neural} {Information} {Processing} {Systems}}, {NIPS}'17, pages 4768--4777,
  Red Hook, NY, USA, December 2017. Curran Associates Inc.
\newblock ISBN 978-1-5108-6096-4.

\end{thebibliography}
\appendix
\section*{Appendix}

\section{Limitations and Ethical Concerns}
\paragraph{Limitations}
Our neural models were trained on one NVIDIA P4 GPU with 8GB RAM, where the maximum training time was 10 minutes for 2,418 conversations for 5 epochs. 
Local non-profits (with scarce funding) may not have access to such GPUs.  
A more optimized training and inference approach will be required to implement this for such use cases.

We assumed that community needs can be classified into categories given by assessments done in the past. However, these categories are by no means exhaustive. 
But linguistically, further work is required into how people approach a conversation regarding needs and we need to deconstruct such an approach computationally to provide a more robust solution. 
Since our corpus in Reddit is already in the form of discussions such deconstruction was not necessary - but our future work will aim to broaden the modalities of conversations to other aspects such as focused-group discussions, one-on-one conversations, town hall meetings, etc. We deconstruct each category of these conversations through multiple methods. However, our approach is not an exhaustive list of all possible ways such analysis can be done.

\icwsm{The Reddit communities represent a narrow demographic that does not capture the full diversity of opinions and people present in their respective communities. Reddit users tend to be younger and more technologically inclined, and hence not fully representative of the population.
The voices captured are limited to those actively engaged in these online spaces which may result in the needs and perspectives of marginalized communities being underrepresented. Furthermore, conversations are shaped by what issues and strengths users are willing and able to discuss online, which may introduce bias and under-representation of sensitive topics compared to a random sampling of the population.}

\paragraph{Ethical Concerns}
All methods in this paper are evaluated on our CNA dataset, but we believe that the linguistic understanding of needs and assets can be applied to similar cases. Furthermore, our dataset was built using only publicly available conversations that any user can view (without requiring to log in to Reddit) and was collected using their public-facing API. The dataset will be released publicly upon the acceptance and publication of this paper and according to the Findable, Accessible, Interoperable, and Reusable (FAIR) principles.
However, there are several ethical concerns when working with such public social media data. It is important to note that the users did not consent to any research studies, and this is an issue similar to a larger issue of using publicly available social media data in research~\cite{giorgi_author_2023}.

\icwsm{
\paragraph{Privacy Concerns} In addressing specific concerns related to the ethical considerations in working with public social media data, it is crucial to acknowledge the unique characteristics of Reddit posts and comments. Reddit inherently maintains a level of user anonymity, where individuals can engage in discussions without revealing their real identities. Furthermore, the entities and locations mentioned in the posts are treated as public information, as they pertain to community-related discussions. The individuals referred to in the comments are public figures, such as community service providers or local politicians, who actively contribute to the community or public sphere.}

\icwsm{To ensure privacy and adhere to ethical standards, we implemented a rigorous process during the dataset creation. A dedicated graduate student researcher meticulously reviewed the entire CNA dataset. The primary objective of this review was to confirm that no personal information leakage occurred and that the dataset exclusively contained the names of community-related individuals. This thorough examination aimed to mitigate any potential risks associated with privacy concerns and reinforce our commitment to handling public social media data responsibly. We recognize the importance of maintaining the integrity of the data and are dedicated to transparently documenting and addressing these ethical considerations in our research. The final dataset will be made publicly available upon the publication of the paper.}

\begin{figure}[h]
    \centering
    \includegraphics[width=0.8\linewidth]{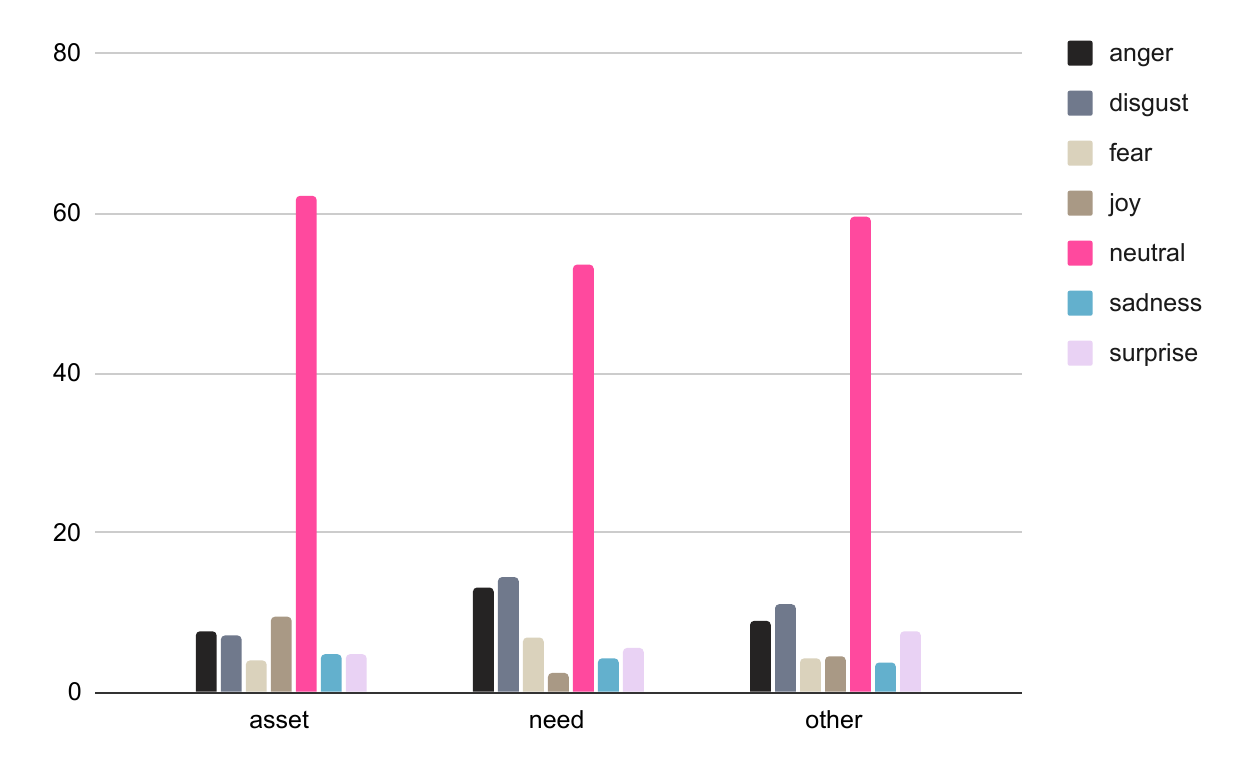}
    \caption{The distribution of emotions for \need, \asset, and other classes in the dataset. The $y$ axis displays the percentage of labels that were assigned the emotions. This includes the \textit{neutral} emotion that was removed from Figure~\ref{fig:emotional_class} for a cleaner picture.}
    \label{fig:emotional_class_appendix}
\end{figure}
\begin{table}[]
    \centering
    \begin{tabular}{lcc}
    \hline
         \textbf{sentiment} & negative & other \\
         \textbf{ground truth}&   {} & {} \\
         need & 552 & 287 \\
         other & 1018 & 1655 \\
         \hline
         \hline
         \textbf{sentiment} & positive & other \\
         \textbf{ground truth}&   {} & {} \\
         asset & 278 & 198 \\
         other & 348 & 2688 \\
         \hline
    \end{tabular}
    \caption{The contingency table for two chi-squared tests of association: (1) \need~and negative sentiment (top table), and (2) \asset~and positive sentiment (bottom table)}
    \label{tab:my_label}
\end{table}
\appendix
\end{document}